\documentclass[pra,showpacs,preprintnumbers]{revtex4-2}
\usepackage{amsmath}
\usepackage{amssymb} 
\usepackage{graphicx}
\usepackage{bm}
\usepackage{comment}
\usepackage{mathrsfs} 
\usepackage{subeqnarray} 
\usepackage{array} 
\usepackage{subfigure} 
\newcommand{\goodgap}{
 \hspace{\subfigtopskip}%
 \hspace{\subfigbottomskip}}%
\usepackage{hyperref}
\usepackage{color}
\usepackage{fancyhdr}
\usepackage{datetime}
\numberwithin{equation}{section} 
\newcommand{\qbin}[3]{\left[\begin{array}{c}#1 \\ #2 \end{array}\right]_{#3}}
\begin{document}
\title{
Angular momentum distribution in a relativistic configuration: 
Magnetic quantum number analysis
}
\author{Michel Poirier}
\email{michel.poirier@cea.fr}
\affiliation{CEA - Paris-Saclay University, Laboratory ``Interactions, Dynamics, 
and Lasers'', CE Saclay, F-91191 Gif-sur-Yvette, France}
\author{Jean-Christophe Pain}
\email{jean-christophe.pain@cea.fr}
\affiliation{CEA, DAM, DIF, F-91297 Arpajon, France}
\affiliation{Universit\'e Paris-Saclay, CEA, Laboratoire Mati\`ere en Conditions 
 Extr\^emes, 91680 Bruy\`eres-le-Ch\^atel, France}
\date{\today}

\begin{abstract}This paper is devoted to the analysis of the distribution of the 
total magnetic quantum number $M$ in a relativistic subshell with $N$ equivalent 
electrons of momentum $j$. This distribution is analyzed through its cumulants 
and through their generating function, for which an analytical expression is 
provided. This function also allows us to get the values of the cumulants at any 
order. Such values are useful to obtain the moments at various orders. Since 
the cumulants of the distinct subshells are additive this study directly applies 
to any relativistic configuration. Recursion relations on the generating function 
are given. It is shown that the generating function of the magnetic quantum 
number distribution may be expressed as a n-th derivative of a polynomial. This 
leads to recurrence relations for this distribution which are very efficient 
even in the case of large $j$ or $N$. The magnetic quantum number distribution 
is numerically studied using the Gram-Charlier and Edgeworth expansions. The 
inclusion of high-order terms may improve the accuracy of the Gram-Charlier 
representation for instance when a small and a large angular momenta coexist 
in the same configuration. However such series does not exhibit convergence 
when high orders are considered and the account for the first two terms often 
provides a fair approximation of the magnetic quantum number distribution. 
The Edgeworth series offers an interesting alternative though this expansion 
is also divergent and of asymptotic nature.
\end{abstract}

\maketitle

\section{Introduction}

The theoretical study of emission or absorption spectral properties of hot 
plasmas, encountered for instance in stellar physics, inertial-confinement 
fusion, or laser-plasma experiments, implies taking into account complex 
ions, i.e., multi-electron configurations with several open subshells. The 
issue of finding the number of states corresponding to a given $(J,M)$ set 
--- $J$ being the magnitude of the total angular momentum operator and $M$ 
the eigenvalue (in units of $\hbar$) of its projection on the $z$-axis --- 
in the case of a set of indistinguishable particles was first investigated 
by Bethe in 1936 for nuclear systems \cite{BETHE36}. The problem of the 
classification of atomic energy levels is discussed in many textbooks about 
quantum mechanics. The determination of the spectroscopic terms arising in a 
given electronic configuration was addressed by different methods, the first 
one being the so-called vector model \cite{COWAN81}. The properties 
(regularities, trends) of such terms were also investigated \cite{JUDD68}. 
The problem of listing the terms arising in a complex configuration can be 
solved from elementary group theory \cite{BREIT26,CURL60,KARAYIANIS65,%
KATRIEL89,XU06}. Besides, the determination of the number of lines between 
two configurations is of great interest. Using group-theoretical methods, 
Krasnitz obtained a compact formula only in the simple case of configurations 
built with non-equivalent electrons \cite{BAUCHE88}. The statistics of 
electric-dipole (E1) lines was studied by Moszkowski \cite{MOSZKOWSKI60}, 
Bancewicz \cite{BANCEWICZ84}, Bauche and Bauche-Arnoult \cite{BAUCHE87,
BAUCHE90}, and more recently by Gilleron and Pain \cite{GILLERON09}. Such a 
quantity is important for opacity codes, for instance, in order to decide 
whether a transition array can be described statistically or requires a  
detailed-line accounting calculation, relying on the diagonalization of the 
Hamiltonian \cite{PORCHEROT11}. In the same spirit, the statistics of 
electric quadrupole (E2) lines was also investigated \cite{PAIN12}. A 
particular case of fluctuation, the odd-even staggering (i.e., the fact that, 
in an electronic configuration, the number of odd values of $J$ can differ 
from the number of even values of $J$), was studied by Bauche and Coss\'e 
\cite{BAUCHE97} and later revisited using the generating-function technique 
\cite{PAIN13}. 

Except maybe for the odd-even staggering, the knowledge of the moments or 
cumulants can be very useful to build a statistical modeling. This was 
carried out by Bauche et al. \cite{BAUCHE87} for the distributions of energy 
levels and spectroscopic terms in an electronic configuration or for the 
distribution of absorption or emission lines. For instance, following the 
pioneering work of Moszkowski \cite{MOSZKOWSKI62}, the first two moments of 
the line energies weighted by their strengths in a transition array were 
calculated exactly by Bauche et al. \cite{BAUCHE88}. The work on averages of 
products of operators by Ginocchio \cite{GINOCCHIO73} enabled Ku\v{c}as and 
Karazija \cite{KARAZIJA91b,KUCAS95II} to find an algorithm to generate the 
moment of any order and the impact of higher-order moments (without 
calculating them explicitly) was studied recently by Gilleron and Pain 
\cite{GILLERON08}. Kyni\.ene et al. investigated the statistical properties 
of Auger transitions and obtained a fair approximation for the 
number of Auger amplitudes \cite{KYNIENE02}. The authors showed that 
statistical properties of Auger spectra mainly depend on the orbital quantum 
numbers of shells involved in the transitions and that rather large values of 
skewness and excess kurtosis indicate a significant deviation of the 
distribution of Auger amplitudes from the normal distribution. Moreover, the 
generating-function formalism is a powerful tool for tackling the counting 
problems, either for finding analytical expressions, deriving recursion 
relations or performing a statistical analysis. Using such a formalism, we 
recently published explicit and recurrence formulas for the number of 
electronic configurations in an atom \cite{Pain2020}, together with a 
statistical analysis through the computation of cumulants.

The object of this work is to show that similar considerations apply to the 
distribution of the magnetic quantum number in a relativistic configuration. 
The present paper is organized as follows. General formulas for the magnetic 
quantum number distribution $P(M)$ are recalled in section \ref{sec:mag}. The 
generating function of cumulants of this distribution in a single- or 
multiple-subshell configuration is derived in the same section. In section 
\ref{sec:PM_nth_der_recur}, recurrence relations are deduced from the 
expression of the quantum number distribution as a n-th derivative. The 
analytical expression of the cumulants is obtained in section 
\ref{sec:det_cum} and an additional recurrence relation for their generating 
function is provided in section \ref{sec:rec_gen}. An analysis of the 
distribution using Gram-Charlier and Edgeworth series is carried out in 
sections \ref{sec:gram} and \ref{sec:Edgeworth} respectively, and the paper 
ends with instructive general considerations about the distribution $P(M)$.

\section{Characterization of the magnetic quantum number distribution: the 
cumulant generating function}\label{sec:mag}
\subsection{Definitions}
Our main objective is to determine the statistics of the angular quantum 
number $J$. However, due to the fact that the quantum number $J$ is the 
eigenvalue of no simple operator, its mathematical study is tedious. 
Therefore, it is more appropriate to study the distribution of the magnetic 
quantum number $M$, another advantage being that when different subshells are 
present their contributions to $M$ simply add up. The $J$ values can be 
obtained from the $M$ values by means of the method of Condon and Shortley 
\cite{Condon1935}, which enables one to express the number $Q(J)$ of levels 
with angular momentum $J$ in a configuration as
\begin{equation}\label{rec}
Q(J)=\sum_{M=J}^{J+1}(-1)^{J-M}P\left(M\right)
 =P\left(M=J\right)-P\left(M=J+1\right),
\end{equation}
where $P$ represents the distribution of the angular-momentum projection $M$. 
In this work, we consider the case of relativistic configurations, which 
means that individual electrons are labeled by their total angular momentum 
$j$. Pauli principle is fully accounted for. For a configuration $j_1^{N_1}
j_2^{N_2}j_3^{N_3}\cdots j_w^{N_w}$, $P(M)$ is determined through the relation
\begin{equation}\label{eq:convolM}
P_{N_1,N_2,\cdots}\left(M\right)=\left(P_{N_1}\otimes P_{N_2}\otimes P_{N_3}
 \otimes\cdots\otimes P_{N_w}\right)\left(M\right),
\end{equation}
where the distributions are convolved two at a time, which means that
\begin{equation}
\left(P_{N_i}\otimes P_{N_j}\right)\left(M\right)=\sum_{M'=-\infty}^{\infty}
 P_{N_i}\left(M'\right) P_{N_j}\left(M-M'\right).
\end{equation}
Let us consider a system of $N$ identical fermions in a configuration consisting 
of a single orbital of degeneracy $g$, $m_i$ being the angular momentum 
projection of electron state $i$. Two constraints must be satisfied:
\begin{equation}
N=n_1+\cdots+n_g=\sum_{i=1}^gn_i,
\end{equation}
where $n_i$ is the number of electrons in state $i$ and
\begin{equation}
M=n_1m_1+\cdots+n_gm_g=\sum_{i=1}^gn_im_i,
\end{equation}
where $n_i=0$ or 1 $\forall i$. In the particular case of the relativistic 
configuration $j^N$, the maximum total angular momentum is
\begin{equation}
J_{\text{max}}=\sum_{m=j-N+1}^jm=(2j+1-N)N/2.
\end{equation}
As stated in statistical treatises \cite{Stuart1994}, the whole information 
about the distribution $P(M)$ of magnetic quantum number is contained in the 
exponential of the cumulant generating function defined as 
\begin{equation}
 \exp(K(t)) = \left<\exp(tM)\right>
 = \sum_M P(M)e^{tM} \left/ \sum_M P(M) \right.\label{eq:defeKt}
\end{equation}
where $P(M)$ is the number of $N$-electron states such as $m_1+\cdots m_N=M$. 
From the Pauli principle this normalization factor is given by the product of 
simple binomial coefficients
\begin{equation}
\sum_M P(M)=\prod_{s=1}^w\binom{2j_s+1}{N_s}.\label{eq:norm}
\end{equation}

\subsection{Derivation from a recurrence relation in a single-subshell case}

As a first step we consider relativistic configurations containing only one 
subshell symbolically written $j^N$. One may express the population $P(M)$ 
as a multiple-sum over each magnetic level population \cite{GILLERON09}
\begin{equation}
P(M)=\sum_{p_1=0}^1\sum_{p_2=0}^1\cdots\sum_{p_g=0}^1
 \delta\left(M-\sum_{k=1}^gp_km_k\right)\ \delta\left(N-\sum_{k=1}^gp_k\right)
\end{equation}
where $p_k$ is the $k$-state population and $g=2j+1$ the subshell degeneracy.
The Kronecker symbol $\delta_{i,j}$ is written here $\delta(i-j)$ for the 
sake of readability.
Each $p_k$ is either 0 or 1, and the individual magnetic quantum numbers are 
\begin{equation}\label{eq:valm}
m_1=-j, m_2=-j+1,\cdots m_g=j.
\end{equation}
Writing the numerator in (\ref{eq:defeKt}) as 
\begin{equation}\label{eq:defs}
s(N,j,t)=\sum_MP(M)e^{Mt}
\end{equation}
one has
\begin{equation}
s(N,j,t)=\sum_M\sum_{p_1=0}^1\sum_{p_2=0}^1\cdots\sum_{p_g=0}^1 
 \delta\left(M-\sum_{k=1}^{g}p_km_k\right)\ 
 \delta\left(N-\sum_{k=1}^{g}p_k\right)e^{Mt}
\end{equation}
in which the sum over $M$ may be eliminated
\begin{equation}
s(N,j,t)=\sum_{p_1=0}^1\sum_{p_2=0}^1\cdots\sum_{p_g=0}^1 
 \delta\left(N-\sum_{k=1}^g p_k\right) \exp\left(\sum_{k=1}^g p_km_kt\right).
\end{equation}
Isolating in this multiple sum the contributions of the $p_g$ index and then 
the $p_1$ index, one gets
\begin{subequations}\begin{align}
s(N,j,t)=&\sum_{p_1=0}^1\sum_{p_2=0}^1\cdots\sum_{p_{g-1}=0}^1 
 \delta\left(N-\sum_{k=1}^{g-1}p_k\right) \exp\left(\sum_{k=1}^{g-1}p_km_kt\right)
 \nonumber\\
& +e^{m_gt}\sum_{p_1=0}^1\sum_{p_2=0}^1\cdots\sum_{p_{g-1}=0}^1 
  \delta\left(N-1-\sum_{k=1}^{g-1}p_k\right)
  \exp\left(\sum_{k=1}^{g-1}p_km_kt\right)\\
=&\sum_{p_2=0}^1\cdots\sum_{p_{g-1}=0}^1 \delta\left(N-\sum_{k=2}^{g-1}p_k\right)
 \exp\left(\sum_{k=2}^{g-1}p_km_kt\right)\nonumber\\
 &+e^{m_1t}\sum_{p_2=0}^1\cdots\sum_{p_{g-1}=0}^1 
 \delta\left(N-1-\sum_{k=2}^{g-1}p_k\right) \exp\left(\sum_{k=2}^{g-1}p_km_kt\right)
 \nonumber\\
 &+e^{m_gt}\left[\sum_{p_2=0}^1\cdots\sum_{p_{g-1}=0}^1 
 \delta\left(N-1-\sum_{k=2}^{g-1}p_k\right) \exp\left(\sum_{k=2}^{g-1}p_km_kt\right)\right.
 \nonumber\\
 &\left.+e^{m_1t}\sum_{p_2=0}^1\cdots\sum_{p_{g-1}=0}^1 
 \delta\left(N-2-\sum_{k=1}^{g-1}p_k\right) \exp\left(\sum_{k=2}^{g-1}p_km_kt\right)\right]
\end{align}\end{subequations}
where we have used the fact that $p_1, p_g$ are equal to 0 or 1. 
One may easily verify that the multiple sum over $p_2\cdots p_{g-1}$ generates 
the subshell with angular momentum $j-1$ and population $M'=\sum_{k=2}^{g-1}p_k$. 
Using the definitions (\ref{eq:valm}), one gets the recurrence property on the 
generating function 
\begin{equation}
s(N,j,t)=s(N,j-1,t)+2\cosh(jt)s(N-1,j-1,t)+s(N-2,j-1,t).\label{eq:recj}
\end{equation}
The argument consisting in specifying the populations $p_1$ and $p_g$ has 
been used by Talmi \cite{Talmi2005} who obtained a recurrence relation on 
the populations $P(M)$ formally similar to Eq.~(\ref{eq:recj}).
The recurrence relation (\ref{eq:recj}) may be initialized by the $N=1$ value. 
One writes from the definition (\ref{eq:defs})
\begin{equation}
s(1,j,t)=\sum_{m=-j}^{j}e^{mt}=e^{-jt}\frac{e^{(2j+1)t}-1}{e^t-1}
 =\frac{\sinh((2j+1)t/2)}{\sinh(t/2)}.\label{eq:vals1}
\end{equation}
Accordingly on may define the initial values for the $j=1/2$ case
\begin{equation}\label{eq:valsj1}
s(0,1/2,t)=s(2,1/2,t)=1,\quad s(1,1/2,t)=2\cosh(t/2),
\quad s(N,1/2,t)=0\text{ if }N>2.
\end{equation}
In order to derive the general expression of the sum $s(N,j,t)$ we performed 
a series of explicit computations for various $N,j$. This work leads us to 
propose the result
\begin{equation}\label{eq:exps}
s(N,j,t)=\frac{\displaystyle\prod_{p=1}^{N}\sinh((2j+2-p)t/2)}{\displaystyle
 \prod_{p=1}^{N}\sinh(pt/2)}.
\end{equation}
This form agrees with the $N=1$ value (\ref{eq:vals1}), and with the $j=1/2$ 
value (\ref{eq:valsj1}). Its general validity is proved here by recurrence. 
Let us assume the property is true up to angular momentum $j-1$. To complete 
the proof one must compute with the above analytical form the ratio 
\begin{equation}
\rho=\Big(s(N,j-1,t)+2\cosh(jt)s(N-1,j-1,t)+s(N-2,j-1,t)\Big)/s(N,j,t)
\end{equation}
and show that it is equal to 1. The expression (\ref{eq:exps}) leads to 
\begin{subequations}\begin{align}
 \rho&=\frac{\sinh((2j+1-N)u)\sinh((2j-N)u)}{\sinh((2j+1)u)\sinh(2ju)}\nonumber\\
  &\qquad+2\cosh(2ju)\frac{\sinh(Nu)\sinh((2j+1-N)u)}{\sinh((2j+1)u)\sinh(2ju)}
  +\frac{\sinh(Nu)\sinh((N-1)u)}{\sinh((2j+1)u)\sinh(2ju)}\\
  &= \frac{\mathscr{N}}{\sinh((2j+1)u)\sinh(2ju)}
\end{align}\end{subequations}
with $u=t/2$ and the numerator
\begin{multline}
\mathscr{N}=\sinh((2j+1-N)u)\sinh((2j-N)u)+2\cosh(2ju)\sinh(Nu)\sinh((2j+1-N)u)\\
  \quad+\sinh(Nu)\sinh((N-1)u).
\end{multline}
Using some elementary trigonometric formulas one easily verifies that 
\begin{equation}\mathscr{N}=\sinh((2j+1)u)\sinh(2ju)\end{equation}
so that $\rho=1$. This completes the proof of (\ref{eq:exps}) by recurrence. 
An alternate derivation based on term counting is briefly mentioned in 
Appendix \ref{sec:relsNsN-1}. Another useful property on the sum $s$ is 
\begin{equation}
s(N,j,t)=\frac{\sinh((2j+2-N)t/2)}{\sinh(Nt/2)}s(N-1,j,t)\label{eq:relSNsN-1}
\end{equation}
from which one can conventionally define
\begin{equation}s(0,j,t)=1\label{eq:s0}\end{equation}whatever $j$.

\subsection{Case of several subshells}
As deduced from a well-known property of the Laplace transform, since the 
distribution $P(M)$ is obtained from the convolution of the distributions of 
every subshell (\ref{eq:convolM}), the Laplace transform for the most general 
relativistic configuration $\sum_M P(M)e^{Mt}$ will be given by the 
\emph{product} of the individual Laplace transforms. For instance if two 
subshells are involved, the exponential of the cumulant generating function 
is given by
\begin{equation}
  e^{K(t)}=\sum_M \sum_{M_1}P_1(M_1)P_2(M-M_1)e^{Mt} 
   \left/ \sum_M \sum_{M_1}P_1(M_1)P_2(M-M_1)\right.
\end{equation}
and since the sums in the numerator and the denominator are the products 
of the individual subshell contributions one easily checks that 
\begin{equation}e^{K(t)}=e^{K_1(t)}e^{K_2(t)}.\end{equation}
In other words, using the analytical form (\ref{eq:exps}) one gets 
for the configuration $j_1^{N_1}...j_w^{N_w}$
\begin{equation}
 e^{K(t)}=\left.\prod_{i=1}^{w}
  \frac{\displaystyle\prod_{p_i=1}^{N_i}\sinh((2j_i+2-p_i)t/2)}{\displaystyle
 \prod_{p_i=1}^{N}\sinh(p_it/2)}\right/
 \prod_{i=1}^{w}\binom{2j_i+1}{N_i}.
\end{equation}
Accordingly, the cumulant generating function $K(t)$ will be given by the 
\emph{sum} of each subshell cumulant generating function.

\section{Expression of the quantum number distribution as a n-th derivative; 
application to recurrence relations}\label{sec:PM_nth_der_recur}
We consider here the case of a configuration made of a single subshell $j^N$. 
The above expression (\ref{eq:exps}) for the exponential of the cumulant 
generating function may be reformulated slightly differently. Defining 
$z=e^t$, the product of hyperbolic sines may be rewritten after simple 
transformations as
\begin{equation}\label{eq:genfnPz}
\sum_M P(M)z^M =z^{-J_\text{max}}\prod_{p=1}^N\frac{z^{2j+2-p}-1}{z^p-1}
\end{equation}
where $J_\text{max}=N(2j+1-N)/2$ is the maximum total angular momentum as 
defined previously. Knowing that
\begin{equation}n=M+J_\text{max}\end{equation}
is an integer varying from 0 to $2J_\text{max}$, one may express $P(M)$ as 
a n-th derivative of the function
\begin{equation}
\mathscr{F}(j,N;z)=\sum_{n=0}^{2J_\text{max}} P(n-J_\text{max})z^n =
 \prod_{p=1}^N\frac{z^{2j+2-p}-1}{z^p-1}.\label{eq:FjNz}
\end{equation}
The $\mathscr{F}(j,N;z)$ function is also known in numerical analysis as the 
Gaussian binomial coefficient or $q$-binomial coefficient \cite{ANDREWS1984}. 
Using standard notation, one has
\begin{equation}\mathscr{F}(j,N;z)=\qbin{2j+1}{N}{z}.\end{equation}
From well-known Pascal-like relations on these polynomials, two recurrence 
relations on the $P(M)$ can be deduced, as shown in Appendix 
\ref{sec:rec_Gaussbin}.

One may also use the expansion (\ref{eq:genfnPz}) to get an expression of the 
$P(M)$ values as an integral. Namely, with $z=e^{2it}$, this expansion can 
be rewritten as
\begin{equation}
\prod_{q=1}^N \frac{\sin((2j+2-q)t)}{\sin(qt)} =  e^{-2iJ_\text{max}t} 
  \sum_{n=0}^{2J_\text{max}}P(n-J_\text{max})e^{2int}
=\sum_{M}P(M)e^{2iMt}.\label{eq:fgen_sin}
\end{equation}
After multiplication by $e^{-2iMt}$ and integration over $t$ on the $[-\pi,\pi]$ 
interval, one gets, accounting for the parity of the above expression
\begin{equation}
P(M)=\frac{1}{2\pi}\int_{-\pi}^{\pi}dt\:e^{-2iMt}
 \prod_{q=1}^N \frac{\sin((2j+2-q)t)}{\sin(qt)}
  =\frac{1}{\pi}\int_{0}^{\pi}dt\:\cos(2Mt)
   \prod_{q=1}^N \frac{\sin((2j+2-q)t)}{\sin(qt)}.
\end{equation}
The above written integrand exhibits a sharp peak close to $t=0$ and
this may be used to derive an approximate value of $P(M)$ using the 
saddle-point method.

Identifying the expansion (\ref{eq:genfnPz}) as a Taylor expansion at $z=0$, 
one gets
\begin{equation}\label{eq:PM_derivn}
P(n-J_\text{max}) = \frac{1}{n!}\left.\frac{d^n}{dz^n}\left(\frac{
 \left(z^{2j+1}-1\right)\left(z^{2j}-1\right)\cdots\left(z^{2j+2-N}-1\right)
 }{\left(z^{N}-1\right)\left(z^{N-1}-1\right)\cdots\left(z-1\right)}
\right) \right|_{z=0}
\end{equation}
which amounts to evaluate the derivative of a rational fraction. One may 
transform the n-th derivative (\ref{eq:PM_derivn}) with the Leibniz rule. 
However while the q-th-derivative at $z=0$ of $z^{2j+2-p}-1$ is elementary 
since equal to $q!\delta(q-(2j+2-p))$, the q-th derivative of $1/(z^p-1)$ is 
nonzero whatever $q$. Therefore the above n-th derivative can be expressed 
via the Leibniz rule as a multiple sum of limited usefulness.

Of course for given $j$ and $N$ a direct analytical computation is tractable. 
For instance if $j=1/2, N=1$
\begin{equation}\mathscr{F}(1/2,1;z)= \frac{z^2-1}{z-1} = 1+z\end{equation}
for which the 0-th and first order derivatives in $z=0$ are 1, so that 
$P(1/2)=P(-1/2)=1$. Accordingly if $j=3/2, N=2$ 
\begin{equation}
\mathscr{F}(3/2,2;z)= \frac{(z^4-1)(z^3-1)}{(z^2-1)(z-1)} 
 = (1+z^2)(1+z+z^2)=1+z+2z^2+z^3+z^4\label{eq:F3h2}
\end{equation}
and the derivatives from order 0 to 4 provide $P(\pm2)=1, P(\pm1)=1, P(0)=2$.
However obtaining an analytical formula valid for any $j$ and $N$ from formula 
(\ref{eq:PM_derivn}) is not straightforward.

Moreover, the identity (\ref{eq:FjNz}) allows the derivation of a recurrence 
property on $N$. The relation 
\begin{equation}
 \mathscr{F}(j,N;z)(z^N-1)=\mathscr{F}(j,N-1;z)(z^{2j+2-N}-1)
\end{equation}
implies, after $n$ derivations with respect to $z$ and use of the Leibniz rule 
\begin{equation}
\sum_{s} \binom{n}{s}\mathscr{F}^{(s)}(j,N;z)(z^N-1)^{(n-s)} = 
 \sum_{t} \binom{n}{t}\mathscr{F}^{(t)}(j,N-1;z)(z^{2j+2-N}-1)^{(n-t)}
\end{equation}
where $f^{(n)}(z)$ is the n-th derivative of $f(z)$ with respect to $z$. 
The above derivatives at $z=0$ are fairly simple. Namely one has
\begin{multline}
 \sum_{s} \binom{n}{s} s!P(s-J_\text{max}(N);j,N)
  \left(N!\delta_{N,n-s}-\delta_{0,n-s}\right)\\
 = \sum_{t} \binom{n}{t} t!P(t-J_\text{max}(N-1);j,N-1)
  \left((2j+2-N)!\delta_{2j+2-N,n-t}-\delta_{0,n-t}\right)
\end{multline}
which provides after basic simplifications the relation between the $P(M;j,N)$ 
and the $P(M';j,N-1)$
\begin{multline}\label{eq:recPM_overN}
P(n-N-N(2j+1-N)/2;j,N) - P(n-N(2j+1-N)/2;j,N) \\
 = P(n-2j-2+N-(N-1)(2j+2-N)/2;j,N-1) -  P(n-(N-1)(2j+2-N)/2;j,N-1).
\end{multline}
With the definition 
\begin{equation}\label{eq:Pscr_def}
 \mathscr{P}_{j,N}(n)=P(n-N(2j+1-N)/2;j,N)
\end{equation}
with $n$ integer in the range $0\le n\le N(2j+1-N)$, one gets the more 
compact formula
\begin{equation}\label{eq:recOPM_overN}
\mathscr{P}_{j,N}(n)=\mathscr{P}_{j,N-1}(n)
 -\mathscr{P}_{j,N-1}(n-2j-2+N)+\mathscr{P}_{j,N}(n-N)
\end{equation}
This relation proves to be very efficient in determining $P(M;j,N)$ 
whatever $j$ and $N$, since the first distribution is elementary
\begin{equation}
 P(M;j,1)=\mathscr{P}_{j,1}(M+j)=1\text{\quad if }-j\le M\le+j.
\end{equation}
For the first $n$ values ($0 \le n <N$), the first member of the recurrence 
(\ref{eq:recPM_overN}) is reduced to the second term 
$P(n-N(2j+1-N)/2;j,N)$. The same behavior occurs for each $n$ below $N$. 
For larger $n$, in the difference $P(n-N-N(2j+1-N)/2;j,N) - 
P(n-N(2j+1-N)/2;j,N)$ the first $P(M)$ has already been computed, which 
defines the population $P(n-N-N(2j+1-N)/2;j,N)$ since the $P(M;j,N-1)$ are 
assumed to be known. 

The identity (\ref{eq:FjNz}) may be used by varying $j$ too. Explicitly
\begin{equation}
\mathscr{F}(j,N;z)(z^{2j-N}-1)(z^{2j+1-N}-1)=
 \mathscr{F}(j-1,N;z)(z^{2j+1}-1)(z^{2j}-1)
\end{equation}
from which one gets after $n$ derivations in $z=0$
\begin{multline}
  \sum_s \binom{n}{s} s!P(s-J_\text{max};j,N)\\
  \times\left[(4j+1-2N)!\delta_{4j+1-2N,n-s}-(2j+1-N)!\delta_{2j+1-N,n-s}
  -(2j-N)!\delta_{2j-N,n-s}+\delta_{0,n-s}\right]\\
  =\sum_t \binom{n}{t} t!P(t-J'_\text{max};j-1,N)
  \left((4j+1)!\delta_{4j+1,n-t}-(2j+1)!\delta_{2j+1,n-t}
  -(2j)!\delta_{2j,n-t}+\delta_{0,n-t}\right)
\end{multline}
with $J_\text{max}=N(2j+1-N)/2$, $J'_\text{max}=N(2j-1-N)/2$. After some 
basic simplifications, one gets the relation involving four $P$ for each $j$ 
value, using the notation (\ref{eq:Pscr_def})
\begin{multline} 
 \mathscr{P}_{j,N}(n-4j-1+2N) -\mathscr{P}_{j,N}(n-2j-1+N)
  -\mathscr{P}_{j,N}(n-2j+N) +\mathscr{P}_{j,N}(n) \\
 = \mathscr{P}_{j-1,N}(n-4j-1) -\mathscr{P}_{j-1,N}(n-2j-1)
  -\mathscr{P}_{j-1,N}(n-2j) +\mathscr{P}_{j-1,N}(n)
\end{multline}
which is less tractable than the recurrence on $N$ (\ref{eq:recOPM_overN}). 
A better option is to allow $j$ to vary by 1/2 instead of 1 and to deal with 
$\mathscr{F}(j,N;z)$ with \textit{integer} $j$ as intermediate calculation 
values without physical meaning. From
\begin{equation}
\mathscr{F}(j,N;z)(z^{2j+1-N}-1)=\mathscr{F}(j-1/2,N;z)(z^{2j+1}-1)
\end{equation}
one gets after multiple derivation in $z=0$
\begin{multline}
 \sum_{s} \binom{n}{s} s!P(s-N(2j+1-N)/2;j,N)
  \left((2j+1-N)!\delta_{2j+1-N,n-s}-\delta_{0,n-s}\right)\\
 = \sum_{t} \binom{n}{t} t!P(t-N(2j-N)/2;j-1/2,N)
  \left((2j+1)!\delta_{2j+1,n-t}-\delta_{0,n-t}\right)
\end{multline}
from which, using the above notation (\ref{eq:Pscr_def})
\begin{equation}\label{eq:recOPM_overj}
 \mathscr{P}_{j,N}(n) = \mathscr{P}_{j-1/2,N}(n) 
 -\mathscr{P}_{j-1/2,N}(n-2j-1) +\mathscr{P}_{j,N}(n-2j-1+N).
\end{equation}

In practical cases, if one has to compute the distribution $P(M)$ for a very 
large $j$ and moderate $N$ the recurrence on $N$ (\ref{eq:recOPM_overN}) will 
be faster. In the opposite situation the recurrence on $j$ 
(\ref{eq:recOPM_overj}) will perform better. 

These properties are interesting alternatives to the method previously 
proposed by Gilleron and Pain \cite{GILLERON09}. To this respect we may 
estimate the number of operations needed to obtain the whole set of 
$P(M)$ values in a $j^N$ relativistic subshell. The brute force technique 
consists in evaluating all the 
\begin{equation}\label{eq:nop_bf}N_\text{bf}=\binom{2j+1}{N}\end{equation}
n-tuple elements and compute the sum $\sum_{i=1}^N m_i$ for each of them. 
The much better alternative provided by the recurrence method by Gilleron 
and Pain \cite{GILLERON09} amounts to perform roughly
\begin{equation}\label{eq:nop_recGP}
N_\text{GP}=N(2j+1)\big(N(2j+1-N)+1\big)
\end{equation}
operations. As a third option, the recurrence over $N$ (\ref{eq:recOPM_overN}) 
will be initialized by the $N=0$ value and then applied for every 
$-J_\text{max}(j,\nu)\le M\le J_\text{max}(j,\nu)$ for $1\le \nu\le N$ with 
$J_\text{max}(j,\nu)=\nu(2j+1-\nu)/2$. Since the formula expresses $P(M)$ as a
function of 3 other $P$s, the number of required operations is 
\begin{equation}\label{eq:nop_recN}
N_\text{recN}=3\sum_{\nu=1}^N \left(\nu(2j+1-\nu)+1\right)
 = N\left(3j(N+1)-N^2+4\right).
\end{equation}
This is even an overestimate since in some cases due to selection rules the 
recurrence formula involves less than 3 terms in its second member. Moreover 
the symmetry property $P(-M)=P(M)$ is not used.
Accordingly the recurrence (\ref{eq:recOPM_overj}) will be used initialized 
with the minimum value $j_0=(N-1)/2$. If $i$ represents twice the iterated 
angular momentum, ranging from $N$ to $2j$, the number of operations will be 
\begin{equation}\label{eq:nop_recj}
N_\text{recj}=3\sum_{i=N}^{2j} \left(N(i+1-N)+1\right)
 = \frac{3}{2}(2j+1-N)\left(N(2j+1-N)+N+2\right).
\end{equation}
Some examples for the numbers (\ref{eq:nop_bf},\ref{eq:nop_recGP},%
\ref{eq:nop_recN},\ref{eq:nop_recj}) are given in Table 
\ref{tab:number_opn_recur}, in the case of an half-filled subshell which 
leads to the maximum complexity. It may be noted that the recurrence on $j$ 
(\ref{eq:recOPM_overj}), though using ``unphysical'' quantities, is sometimes 
more efficient than the recurrence on $N$.
\begin{table}[htbp]
\renewcommand{\arraystretch}{1.5}
\begin{tabular}{c *8{>{\raggedleft\arraybackslash}m{1.5cm}}}
\hline\hline
$j$           & 1/2 & 3/2 & 7/2 & 11/2 & 15/2 & 19/2 & 23/2 & 27/2\\
\hline
$N_\text{bf}$   & 2 &  6 &  70 & 924 & 12870 & 184756 & 2704156 & 40116600 \\
$N_\text{GP}$   & 4 & 40 & 544 &2664 &  8320 &  20200 &   41760 &    77224 \\
$N_\text{recN}$ & 6 & 27 & 162 & 501 &  1140 &   2175 &    3702 &     5817 \\
$N_\text{recj}$ & 6 & 24 & 132 & 396 &   888 &   1680 &    2844 &     4452 \\
\hline\hline
\end{tabular}
\caption{Number of operations needed to obtain the $P(M)$ distribution 
for the $j^N$ configuration with $N=j+1/2$, using a brute-force technique 
or recurrence relations. Numbers are given according to formulas 
(\ref{eq:nop_bf},\ref{eq:nop_recGP},\ref{eq:nop_recN},\ref{eq:nop_recj}).
\label{tab:number_opn_recur}}
\end{table}

\section{Determination of the cumulants and moments}\label{sec:det_cum}
\subsection{Analytical form of the cumulants}
According to the definitions (\ref{eq:defeKt}) and the normalization 
(\ref{eq:norm}) the cumulant generating function is
\begin{subequations}\begin{align}\label{eq:defKt}
K(t)&=\log\left(\sum_M P(M)e^{Mt}\left/\binom{2j+1}{N}\right.\right)\\
 &= \sum_{p=1}^N \Big[\log(\sinh((2j+2-p)t/2) -\log(2j+2-p) 
  -\log(\sinh(pt/2)) +\log p\,\Big].
\end{align}\end{subequations}
From the expansion
\begin{equation}\label{eq:devlogsh}
\log(\sinh x) = \log x 
 +\sum_{n=1}^\infty \frac{B_{2n}}{2n}\frac{(2x)^{2n}}{(2n)!}
\end{equation}
where $B_j$ are the Bernoulli numbers \cite{Abramowitz1972}, one gets the 
series expansion for the cumulant generating function 
\begin{equation}
K(t)=\sum_{k=1}^\infty \frac{B_{2k}}{2k}\left[\sum_{p=1}^{N}(2j+2-p)^{2k} 
 - \sum_{p=1}^N p^{2k} \right] \frac{t^{2k}}{(2k)!}.\label{eq:exprcumulgen}
\end{equation}
This expansion allows us to obtain the cumulants $\kappa_n$ defined by 
\cite{Stuart1994}
\begin{equation}
K(t)=\sum_{n=1}^\infty \kappa_n\frac{t^n}{n!}
\end{equation}
where $\kappa_1$ is the distribution average, $\kappa_2$ the variance, 
$\kappa_3$ the asymmetry, $\kappa_4$ the excess kurtosis, etc. Identifying 
this expansion with the analytical form (\ref{eq:exprcumulgen}) one directly 
obtains the even-order cumulants of the $M$ distribution
\begin{equation}
\kappa_{2k}=\frac{B_{2k}}{2k} \left[\sum_{p=1}^{N}(2j+2-p)^{2k} 
 - \sum_{p=1}^N p^{2k} \right]\label{eq:cumul}
\end{equation}
while of course odd-order cumulants vanish. This expression may be rewritten 
\begin{equation}
\kappa_{2n}(j^N)=\frac{B_{2n}}{2n}
 \left[\sum_{p=0}^{2j+1}p^{2n}-\sum_{p=0}^{N}p^{2n}-\sum_{p=0}^{2j+1-N} p^{2n}\right]
\end{equation}
which makes more obvious the invariance of the cumulant under the 
transformation $N\rightarrow 2j+1-N$. Using the relation
\begin{equation}
\sum_{k=0}^{n-1}k^m=\frac{B_{m+1}(n)-B_{m+1}(0)}{m+1},
\end{equation}
where $B_n(x)$ is the $n$-th Bernoulli polynomial \cite{Abramowitz1972}, one gets 
\begin{equation}
\kappa_{2n}(j^N)=\frac{B_{2n}}{2n}
 \frac{\left[B_{2n+1}(2j+2)-B_{2n+1}(N)-B_{2n+1}(2j+2-N)\right]}{2n+1}.
\end{equation}

\subsection{Explicit expressions for the first cumulants}\label{sec:expli_cum}
A careful analysis of the formula (\ref{eq:cumul}) shows that the 
cumulant at order $2k$ may be expressed as a polynomial of order $2k$ in $N$. 
Furthermore, because of the symmetry $P(M)=P(2j+1-N)$, one knows that changing 
$N\rightarrow 2j+1-N$ the cumulant must be invariant. Therefore this cumulant 
must be a polynomial of order $k$ in the variable $N(2j+1-N)$. One defines
\begin{equation}\label{eq:kap2kwithC}
\kappa_{2k}=\sum_{p=1}^{k}C(2k,p)\left[N(2j+1-N)\right]^p.
\end{equation}
The values for $C(2k,p)$ have been computed for $k$ up to 6 with Mathematica 
software using the explicit form (\ref{eq:cumul}). One gets
\begingroup
\allowdisplaybreaks
\begin{subequations}\begin{align}
C(2,1)&=\frac{j+1}{6}\label{eq:variance}\\
C(4,1)&=-\frac{1}{30}(j+1)^2(2j+1)\\
C(4,2)&=\frac{j+1}{60}\\
C(6,1)&=\frac{1}{126}(j+1)^2(2j+1)\left(8j^2+12j+3\right)\\
C(6,2)&=-\frac{1}{252}(j+1)(2j+1)(8j+9)\\
C(6,3)&=\frac{j+1}{126}\\
C(8,1)&=-\frac{1}{90}(j+1)^2(2j+1)\left(24j^4+72j^3+70j^2+24j+3\right)\\
C(8,2)&=\frac{1}{180}(j+1)(2j+1)\left(36j^3+96j^2+76j+15\right)\\
C(8,3)&=-\frac{1}{90}(j+1)(2j+1)(5j+6)\\
C(8,4)&=\frac{j+1}{120}\\
C(10,1)&=\frac{1}{66}(j+1)^2(2j+1)\left(4j^2+6j+1\right)
 \left(32j^4+96j^3+92j^2+30j+5\right)\\
C(10,2)&=-\frac{1}{132}(j+1)(2j+1)\left(256j^5+1072j^4+1648j^3+1112j^2+312j+35\right)\\
C(10,3)&=\frac{1}{66}(j+1)(2j+1)\left(56j^3+156j^2+128j+25\right)\\
C(10,4)&=-\frac{5}{132}(j+1)(2j+1)(4j+5)\\
C(10,5)&=\frac{j+1}{66}\\
C(12,1)&=-\frac{691}{8190}(j+1)^2(2j+1)\nonumber\\ &\quad\times
 \left(256j^8+1536j^7+3712j^6+4608j^5+3160j^4+1272j^3+338j^2+48j+3\right)\\
C(12,2)&=\frac{691}{16380}(j+1)(2j+1)\nonumber\\ &\quad\times
 \left(640j^7+3648j^6+8288j^5+9488j^4+5756j^3+1872j^2+356j+27\right)\\
C(12,3)&=-\frac{691}{8190}(j+1)(2j+1)\left(192j^5+832j^4+1316j^3+900j^2+247j+28\right)\\
C(12,4)&=\frac{691}{32760}(j+1)(2j+1)\left(224j^3+644j^2+542j+105\right)\\
C(12,5)&=-\frac{691}{8190}(j+1)(2j+1)(7j+9)\\
C(12,6)&=\frac{691}{16380}(j+1)
\end{align}\end{subequations}
\endgroup

\subsection{Computation of the distribution moments}
From these expressions one may also derive the even-order moments, i.e., the 
average values inside a relativistic subshell
\begin{equation}\label{eq:defmt}\mu_{2k}=\sum_M M^{2k}P(M)/\sum_M P(M).
\end{equation}
The relation between moments and cumulants, found in textbooks about 
statistics \cite{Stuart1994}, may be written as
\begin{equation}\label{eq:momts_cumults}
\mu_n = \kappa_n + \sum_{m=1}^{n-1}\binom{n-1}{m-1}\kappa_m\mu_{n-m}.
\end{equation}
The expressions for the moments $\mu_{2k}$ are given in the appendix 
\ref{sec:appmomts} for $k$ up to 6.

\section{Another recurrence relation on the generating function}\label{sec:rec_gen}
Another relation between $s_N$ and values for lower $N$ but the same $j$ may 
be obtained considering the explicit sum definition with $m_k$ indices. 
Defining
\begin{equation}S_N(t)=N!s(N,j,t)\end{equation}
one has
\begin{subequations}\begin{align}
S_N(t) &= N!\sum_{m_1<m_2<\cdots<m_N}e^{(m_1+\cdots+m_N)t} = 
 \sum_{\stackrel{m_1\cdots m_N}{\text{all }\ne}} e^{(m_1+\cdots+m_N)t}\\
 &= \sum_{\stackrel{m_1\cdots m_{N-1}}{\text{all }\ne}} 
  e^{(m_1+\cdots+m_{N-1})t}\sum_{m_N}e^{m_Nt}
  -(N-1)\sum_{\stackrel{m_1\cdots m_{N-1}}{\text{all }\ne}}
   e^{(m_1+\cdots+m_{N-2}+2m_{N-1})t}\\
 &= S_{N-1}(t)S_1(t) 
  -(N-1)\sum_{\stackrel{m_1\cdots m_{N-1}}{\text{all }\ne}}
   e^{(m_1+\cdots+m_{N-2}+2m_{N-1})t}
\end{align}
and repeating the process for the sum over $m_1\cdots m_{N-1}$
\begin{align}
S_N(t) &= S_{N-1}(t)S_1(t) 
 -(N-1)\sum_{\stackrel{m_1\cdots m_{N-2}}{\text{all }\ne}}
  e^{(m_1+\cdots+m_{N-2})t}S_1(2t) \nonumber\\
 &\quad+(N-1)(N-2)\sum_{\stackrel{m_1\cdots m_{N-2}}{\text{all }\ne}}
  e^{(m_1+\cdots+m_{N-3}+3m_{N-2})t}\\
 &= S_{N-1}(t)S_1(t)-(N-1)S_{N-2}(t)S_1(2t)+(N-1)(N-2)S_{N-3}(t)S_1(3t)\nonumber\\
  &\quad -(N-1)(N-2)(N-3)
 \sum_{\stackrel{m_1\cdots m_{N-3}}{\text{all }\ne}}
 e^{(m_1+\cdots+m_{N-4}+4m_{N-3})t}.
\end{align}\end{subequations}
One verifies that the $k$-th term in the expansion is $(-1)^{k-1}(N-1)
\cdots(N-k+1)S_{N-k}(t)S_1(kt)$. The recurrence is closed by studying the 
last two-index sum for $k=N-1$. One has
\begin{equation}
 (-1)^{N-2}(N-1)\cdots2\sum_{\stackrel{m_1,m_2}{m_1\ne m_2}}e^{\Big(m_1+(N-1)m_2\Big)t}
  = (-1)^{N-2}\frac{(N-1)!}{1!}\left[S_1(t)S_1((N-1)t)-S_1(Nt)\right].
 \end{equation}
We have thus proven the general formula
\begin{equation}
S_N(t)=S_{N-1}(t)S_1(t)+\sum_{p=2}^{N-1}(-1)^{p-1}\frac{(N-1)!}{(N-p)!}S_{N-p}(t)S_1(pt)
 +(-1)^{N-1}\frac{(N-1)!}{0!}S_1(Nt).\label{eq:devSN}
\end{equation}
This equation may be simplified using the initial value (\ref{eq:s0}) which allows us 
to write the above sum as 
\begin{equation}\label{eq:recSN}
 s(N,j,t)=\frac1N \sum_{p=1}^{N}(-1)^{p-1}s(N-p,j,t)s(1,j,pt).
\end{equation}
From this expression one obtains a recurrence relation on the distribution moments, 
as shown in Appendix \ref{sec:rec_mom}.

\section{Gram-Charlier series}\label{sec:gram}
\subsection{General formulas}
An interesting property of distributions for which the moments or the cumulants 
are known up to a certain order is that they can be approximated by analytical 
forms. The magnetic quantum number distribution in any relativistic configuration 
may be approximated by a Gram-Charlier expansion defined as (see Sec. 6.17 in 
Ref.\cite{Stuart1994})
\begin{equation}\label{eq:Gram-Charlier}
F_\text{GC}(M) = \frac{G}{(2\pi)^{1/2}\sigma}
 \exp\left[-\frac{(M-\left<M\right>)^2}{2\sigma^2}\right] \left[1+\sum_{k\ge 3}
 c_k He_k\left(\frac{M-\left<M\right>}{\sigma}\right)\right]
\end{equation}
in which $M$ is allowed to vary continuously, while the mean value $\left<M\right>$ 
vanishes for symmetry reasons. In the above equation, $He_n$ is the 
Chebyshev-Hermite polynomial \cite{Stuart1994} 
\begin{equation}\label{eq:Hermite_stat}He_k(X) = 
k!\sum_{m=0}^{\lfloor k/2\rfloor}\frac{(-1)^m X^{k-2m}}{2^m m! (k-2m)!}
\end{equation}
and $\lfloor x\rfloor$ is the integer part of $x$. The Gram-Charlier 
coefficients $c_k$ are related to the moments $\mu_k$ --- which 
are here \textit{centred}, i.e, $\mu_1=0$ --- through the relation
\begin{equation}
c_k=\sum_{j=0}^{\lfloor k/2\rfloor}\frac{(-1)^j
 \mu_{k-2j}/\sigma^{k-2j}}{2^j j!(k-2j)!}\label{eq:cGC_vs_momts}
\end{equation}
and from this definition the coefficients $c_1$ and $c_2$ cancel. It is 
interesting to note that Ginocchio and Yen have used a very similar approach 
to model the state density in nuclei \cite{GINOCCHIO1975}. However in the 
case they considered, the asymmetry term $c_3$ was present and the expansion 
was truncated after the fourth term (excess kurtosis). 

For a symmetric distribution considered here, all the odd-order terms 
$c_{2k+1}$ vanish. The coefficient $G$ in Eq. (\ref{eq:Gram-Charlier}) 
is given by the normalization condition
\begin{equation}
G = \int_{-\infty}^{\infty}\!\!dM\; F_\text{GC}(M) =
 \prod_s\binom{2j_s+1}{N_s},
\end{equation}
the average value is 0, and the variance is derived from (\ref{eq:variance})
\begin{equation}
\sigma^2=\frac16\sum_s (j_s+1)N_s(2j_s+1-N_s).
\end{equation}
As shown in Appendix of our previous paper \cite{Pain2020}, one may also 
express the Gram-Charlier coefficients as a function of the cumulants. For 
instance owing to the parity of $P(M)$, one has $c_4=\kappa_4/(4!\sigma^4)$, 
$c_6=\kappa_6/(6!\sigma^6)$, $c_8=\Big(\kappa_4/(4!\sigma^4)\Big)^2/2+
\kappa_8/(8!\sigma^8)$, etc. Since the cumulants $\kappa_{2k}$ are easily 
obtained from their analytical expression for any relativistic configuration
this might look as the preferred method. However in order to get $c_{2k}$ this 
procedure requires to build the various partitions of the integer $k$, 
which becomes tedious when $k$ is large. Therefore we have used the relation 
(\ref{eq:cGC_vs_momts}), the moments at any order being given by formula 
(\ref{eq:momts_cumults}).

The Gram-Charlier expansion truncated for various $k_\text{max}$ has been 
computed and compared to exact values for the $P(M)$ distribution. 
The exact values were obtained exactly from the recursive procedure described 
by Gilleron and Pain \cite{GILLERON09} or from the above recurrence relations 
on $N$ (\ref{eq:recOPM_overN}) or $j$ (\ref{eq:recOPM_overj}). In the following 
subsections, ``GC 1 term'' will refer to the value of this series for 
$k_\text{max}=2$, i.e., the plain Gaussian form, ``GC 2 terms'' is the series 
truncated at $k_\text{max}=4$, i.e., involving the excess kurtosis, etc.

\subsection{Numerical accuracy and convergence considerations}
The accuracy of the Gram-Charlier series (\ref{eq:Gram-Charlier}) is evaluated 
by truncating the series at some maximum $k$. Let us define
\begin{equation}
P_\text{GC}(M;k_\text{max})=\frac{G}{(2\pi)^{1/2}\sigma}
 \exp\left(-M^2/2\sigma^2\right) \left(1+\sum_{k\le k_\text{max}}
 c_k He_k\left(M/\sigma\right)\right).
\end{equation}
We define the global absolute error as
\begin{equation}\label{eq:abserr}
\Delta_\text{abs}(k)= \left[\sum_{M=-J_\text{max}}^{J_\text{max}}
 \Big(P_\text{GC}(M;k)-P(M)\Big)^2/(2J_\text{max}+1)\right]^{1/2}
\end{equation}
and the global relative error 
\begin{equation}\label{eq:relerr}
\Delta_\text{rel}(k)= \left[\sum_{M=-J_\text{max}}^{J_\text{max}}
 \Big(P_\text{GC}(M;k)/P(M)-1\Big)^2/(2J_\text{max}+1)\right]^{1/2}.
\end{equation}
We have computed Gram-Charlier series in a wide range of cases using first a 
fully numerical approach with high floating-point accuracy (Fortran with 
16-byte real numbers, i.e., about 32-digit accuracy), then using formal 
calculation through Mathematica software working with arbitrary precision 
--- the $c_k$ coefficients are indeed rational fractions which can be 
manipulated ``exactly'', the only numerical conversion being done when 
the non-rational exponential and the normalization factors in 
Eq.~(\ref{eq:Gram-Charlier}) are computed.
We observed that these two approaches provide very different results 
when high order terms are computed. Indeed, while the moments $\mu_{2k}$ 
are all positive, the coefficients (\ref{eq:cGC_vs_momts}) of this 
series involve a sum with alternating signs. The definition of $c_k$ 
as a function of the cumulants \cite{Pain2020} only involves positive 
coefficients, but the cumulants themselves are of alternate signs. We 
could numerically check that when considering very large $k$ the series 
$c_k$ indeed tends to 0 but the Fortran computation provides values larger 
by order of magnitudes than the Mathematica computation. This numerical 
divergence may appear for $k$ not greater than 50. For instance in the 
case illustrated by Fig. \ref{fig:GC_j7N4_distribM_err}, we noticed 
a very strong divergence of the 16-byte computation for $k\simeq40$. 
As a consequence, when numerical instabilities were observed, we have 
monitored the computation accuracy by comparing to arbitrary precision 
results. This leads us to realize that some of the ``divergences'' observed 
in our previous work \cite{Pain2020} were of numerical nature. However one 
must keep in mind that due to the strong compensation occurring in the 
Gram-Charlier coefficient computation any numerical approach will encounter 
this loss of accuracy when high enough orders are reached. General 
considerations about the Gram-Charlier series convergence will be provided at 
the end of this Section. 

\subsection{Example of small \textit{j} and small \textit{N}}
\begin{figure}[htbp]
\centering
\subfigure[$M$ distribution: exact and Gram-Charlier approximation. 
 The Gram-Charlier expansion includes 1, 2, 3, 4, 8, 16 or 32 terms in 
 the sum (\ref{eq:Gram-Charlier}).
]{\label{fig:PM_GC_j7N4}
\includegraphics[scale=0.3]{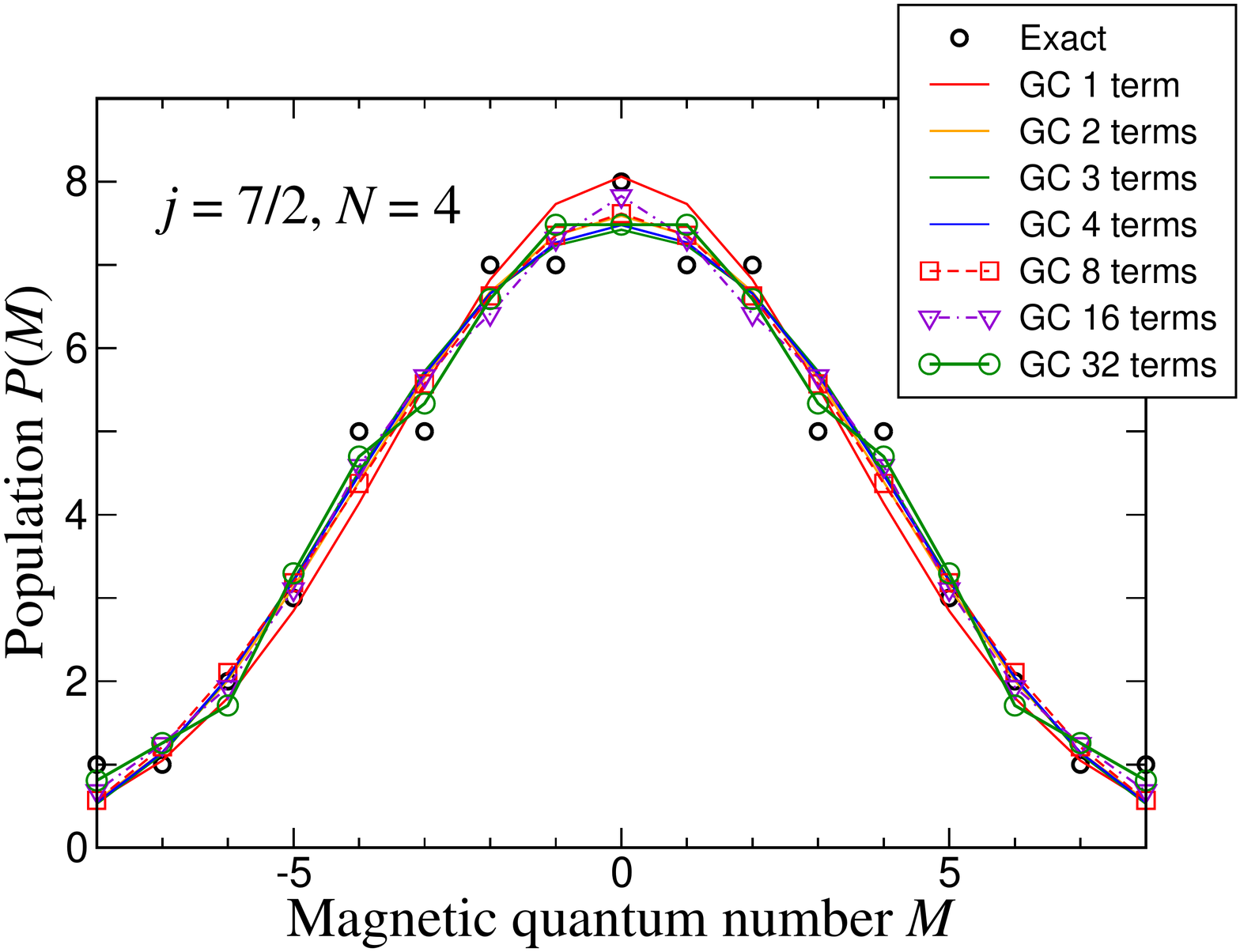}
}
\hspace{0.1cm}
\subfigure[Evaluation of the Gram-Charlier approximation. 
The average absolute and relative errors are defined in main text. 
The error at $M_0=J_\text{max}$ or $M_0=0$ is the value of the 
difference $P_\text{GC}(M_0;k_\text{max})-P(M_0)$ plotted 
as a function of the half truncation index $k_\text{max}/2$.
]{\label{fig:error_GC_j7N4}
\includegraphics[scale=0.3]{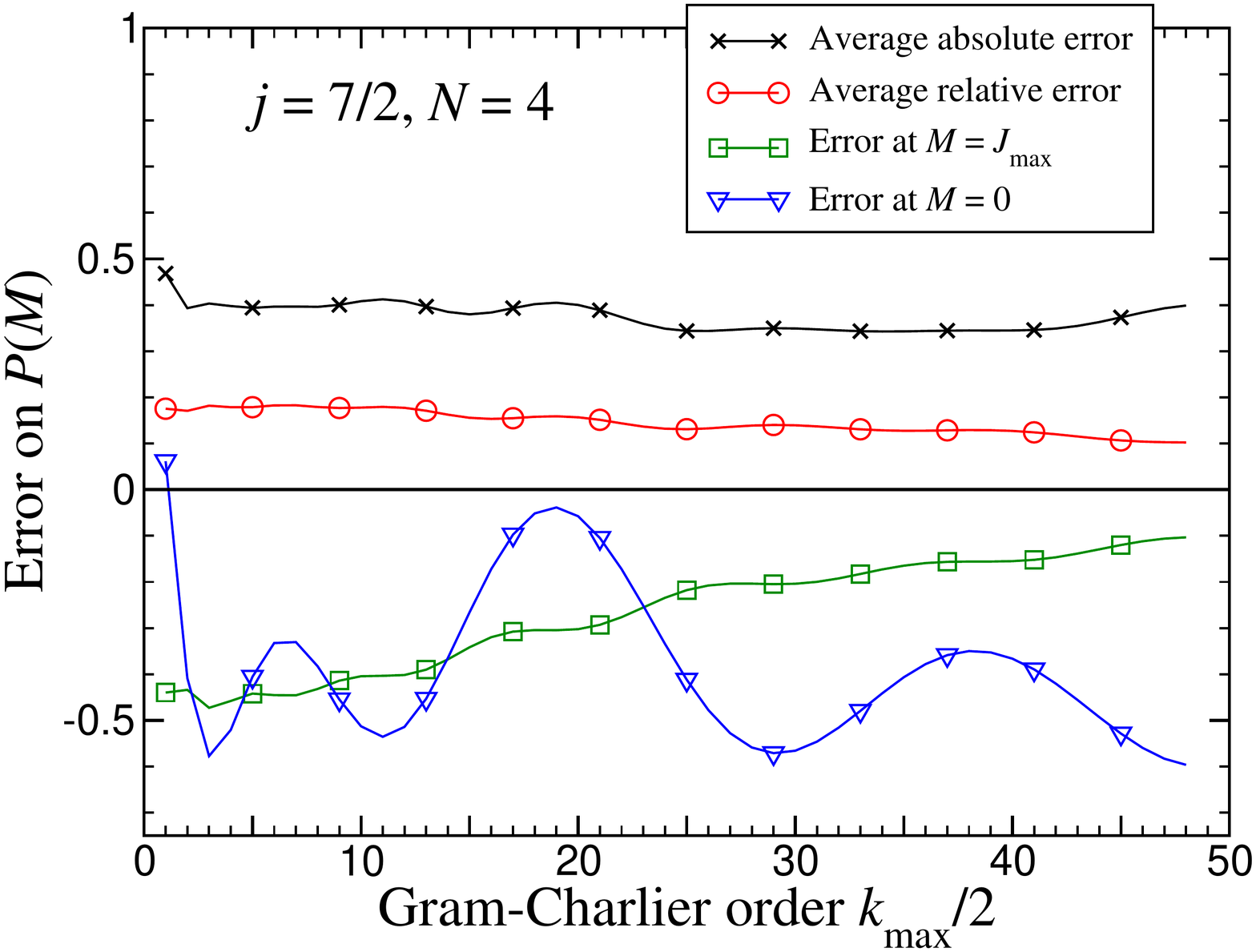}
}
\caption{Exact and Gram-Charlier approximation for the magnetic quantum 
number $P(M)$ in the relativistic configuration $j=7/2$, $N=4$ (e.g., 
$4g_{7/2}^4$). The one-term computation is the Gaussian form, the 2-term form 
includes the $c_4$ contribution, i.e., the excess kurtosis, the 3-term form 
includes $c_4$ and $c_6$, etc. 
\label{fig:GC_j7N4_distribM_err}}
\end{figure}

As a first example we compare on Fig.~\ref{fig:PM_GC_j7N4} the exact $P(M)$ 
distribution and its Gram-Charlier approximation in the case $j=7/2$ and 
$N=4$, for which $J_\text{max}=8$. 
Configurations with similar $j$ and $N$ are quite common in plasma 
spectroscopy, for instance in the context of source design for 
nanolithography \cite{SHEVELKO1998}. In the $j=7/2$ and $N=4$ case, the 
distribution $P(M)$ exhibits plateaus at $P=1$, 5 and 7, and can hardly be 
described by a Gaussian form with great accuracy. Nevertheless except for 
$M=\pm8$ the relative accuracy is about 10\%. One notes on this figure that 
including as many as 32 terms in the Gram-Charlier series does not 
significantly improves the agreement. To get a more quantitative description 
we have plotted in Fig.~\ref{fig:error_GC_j7N4} various accuracy estimates 
for the Gram-Charlier expansion. The absolute error defined by 
Eq. (\ref{eq:abserr}) and the relative error from Eq. (\ref{eq:relerr}) 
are plotted as a function of the half truncation index $k/2$. On this figure 
we have also plotted the errors $P_\text{GC}(M;k)-P(M)$ for $M=0$ and 
$M=J_\text{max}$. One observes that the relative error (\ref{eq:relerr}) 
stays roughly constant at 20\% for $k<32$.  
One notes that absolute error is almost constant for large $k$, while 
the error at $M=\pm J_\text{max}$ slowly decreases with $k$. It turns out 
that the residual error comes from the $M$values close to 0. As seen from 
the definition (\ref{eq:relerr}) the relative error is mostly sensitive to 
the $M$ values where $P(M)$ is small, i.e., $|M|\simeq J_\text{max}$, and 
accordingly this error slowly drops when $k_\text{max}$ increases.

\begin{figure}[htbp]
 \centering
 \subfigure[$M$ distribution]{\label{fig:PM_GC_j5n2j37}
\includegraphics[scale=0.3]{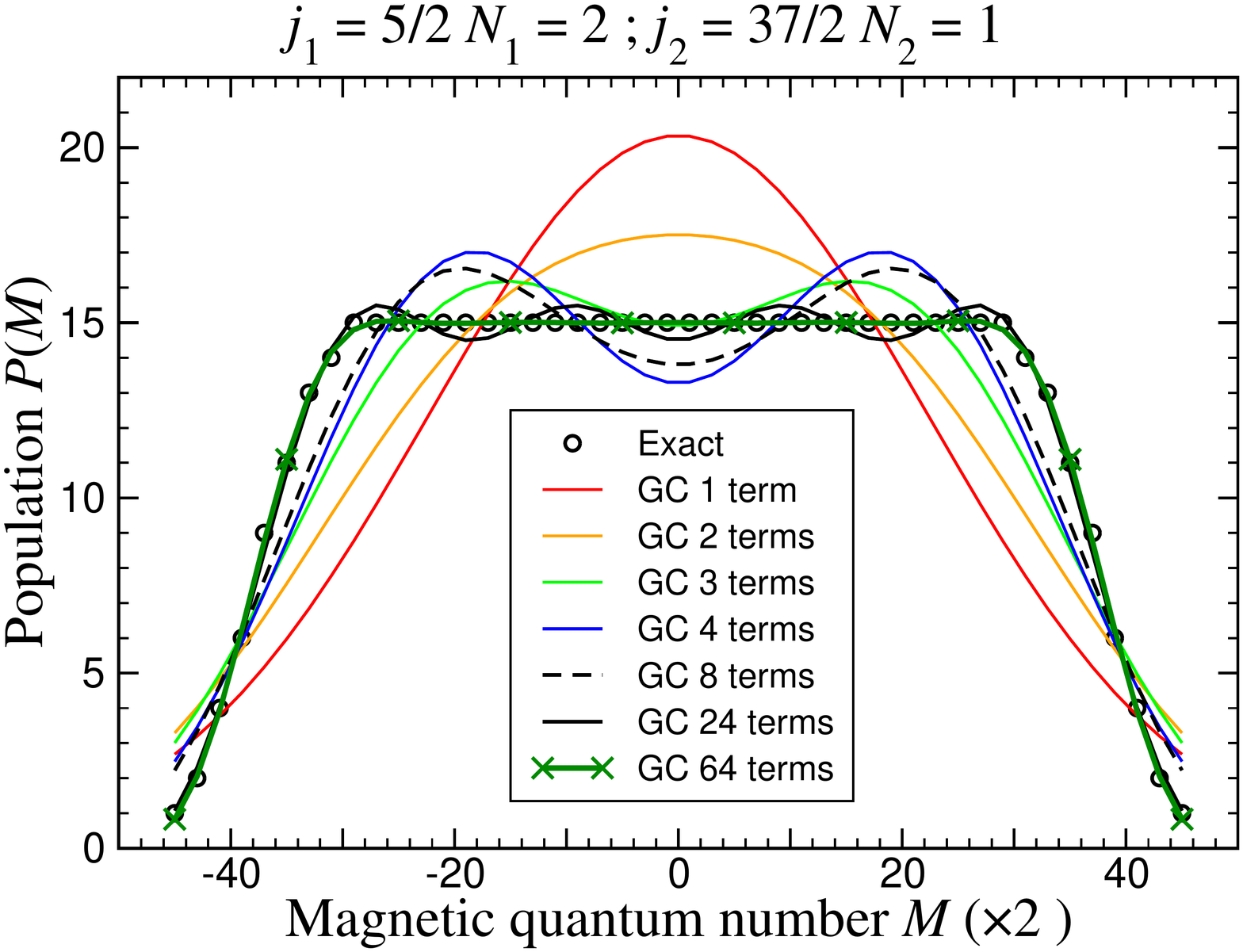}
}\goodgap%
\subfigure[Accuracy of Gram-Charlier approximation as a function of the 
``half-truncation order'' $k/2$. See main text for the definitions.
]{\label{fig:error_GC_j5n2j37}
\includegraphics[scale=0.3]{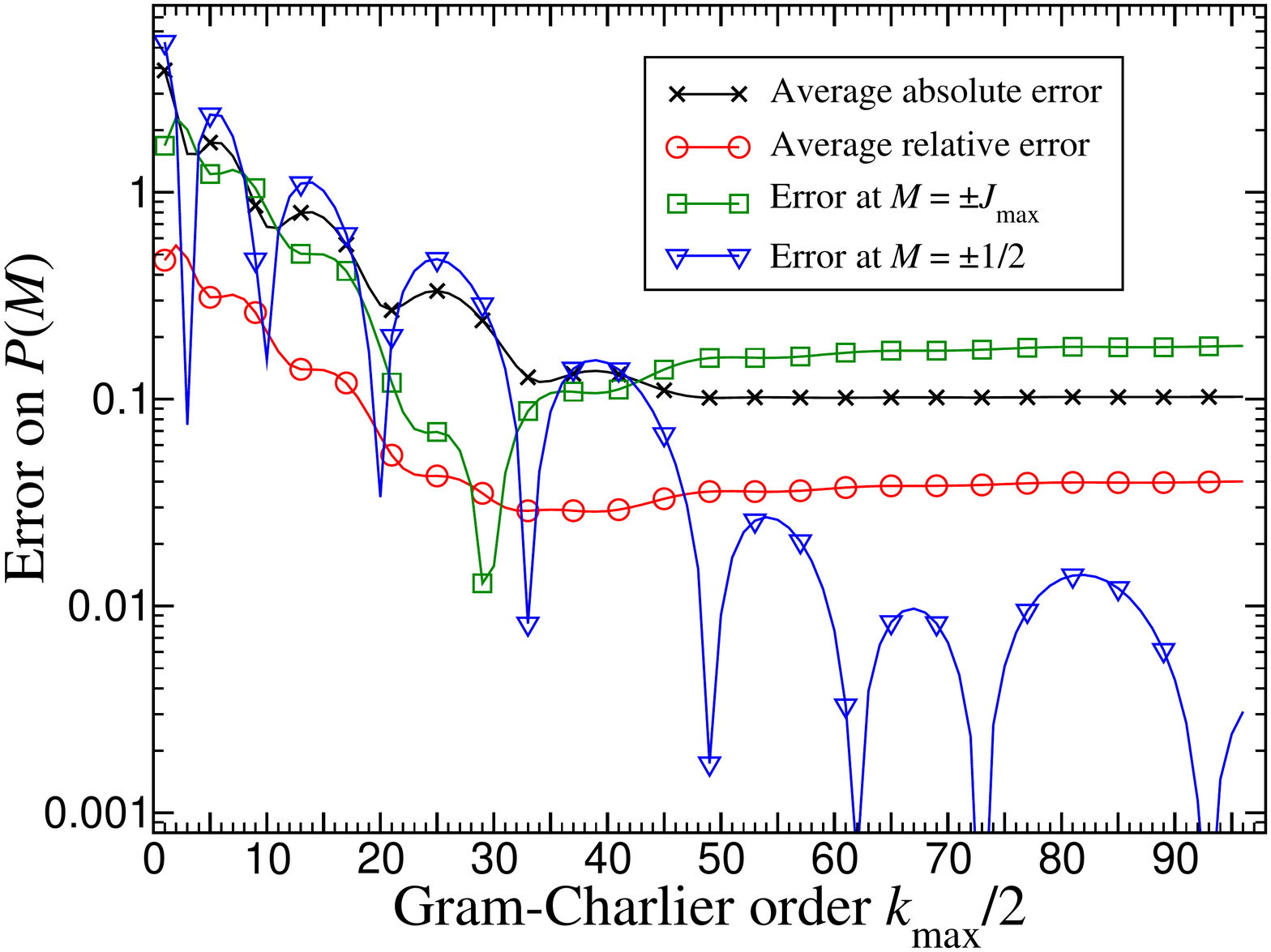}
}
 \caption{Exact and Gram-Charlier approximations for the magnetic quantum 
number $P(M)$ in the relativistic configuration $j_1=5/2$ $N_1=2, j_2=37/2, 
N_2=1$. 
\label{fig:GC_j5N2j37_distribM_err}}
\end{figure}

Another interesting example is provided by $P(M)$ distributions exhibiting 
a wide plateau around $M=0$, which occur in configurations containing both 
high and low $j$ values. Configurations involving high-$j$ spectators are 
created for instance by electron capture into high-lying Rydberg states in 
collisions between multiply charged ions and light target gases 
\cite{HVELPLUND1981}. Let us consider the configuration $j_1=5/2, N_1=2, 
j_2=37/2, N_2=1$ which is analogous to the case considered in 
Ref.~\cite{GILLERON09}.
The magnetic quantum number distribution for this case is plotted in 
Fig.~\ref{fig:PM_GC_j5n2j37}. One notes that the first orders of the 
Gram-Charlier expansion provide a poor representation of the wide plateau 
extending from $M=-29/2$ to $M=29/2$. The quality of this approximation 
slowly improves with $k_\text{max}$, but obtaining a good agreement with 
the exact $P(M)$ distribution requires large $k_\text{max}$ values.
The evolution of the accuracy with the truncation index in the Gram-Charlier 
series is quantitatively analyzed on Fig.~\ref{fig:GC_j5N2j37_distribM_err}. 
It appears 
that all the $P(M)$ values, including those for $M\simeq0$ and $M\simeq
\pm J_\text{max}$ are correctly described for a cut-off $k_\text{max} 
/2\simeq40$. The average absolute error is then $0.28$, the average 
relative error is $0.066$, the error at $M=\pm J_\text{max}$ is 
$0.18$ and the error at $M=\pm1/2$ is $-0.035$, which means that the 
relative error $|P_\text{GC}(M;k)/P(M)-1|$ is below 15\% for any $M$. 
Above this $k$ value, adding more terms slightly improves the accuracy 
in the $M\simeq0$ region, while the larger $|M|$ values are almost 
insensitive to these high-order terms. Though we did not develop a 
rigorous mathematical analysis, it appears that the Gram-Charlier series 
provides an asymptotic-type convergence: for a large range of 
$k_\text{max}$ values, the absolute error levels off at 0.28, and for 
very large truncation index, a divergence is expected.

\subsection{Example of large \textit{j} and large \textit{N}}
\begin{figure}[htbp]
 \centering
 \subfigure[$M$ distribution. For $k_\text{max}\ge4$ the plots are almost 
indistinguishable at the drawing accuracy.]
{\label{fig:PM_GC_j15N8}
\includegraphics[scale=0.3]{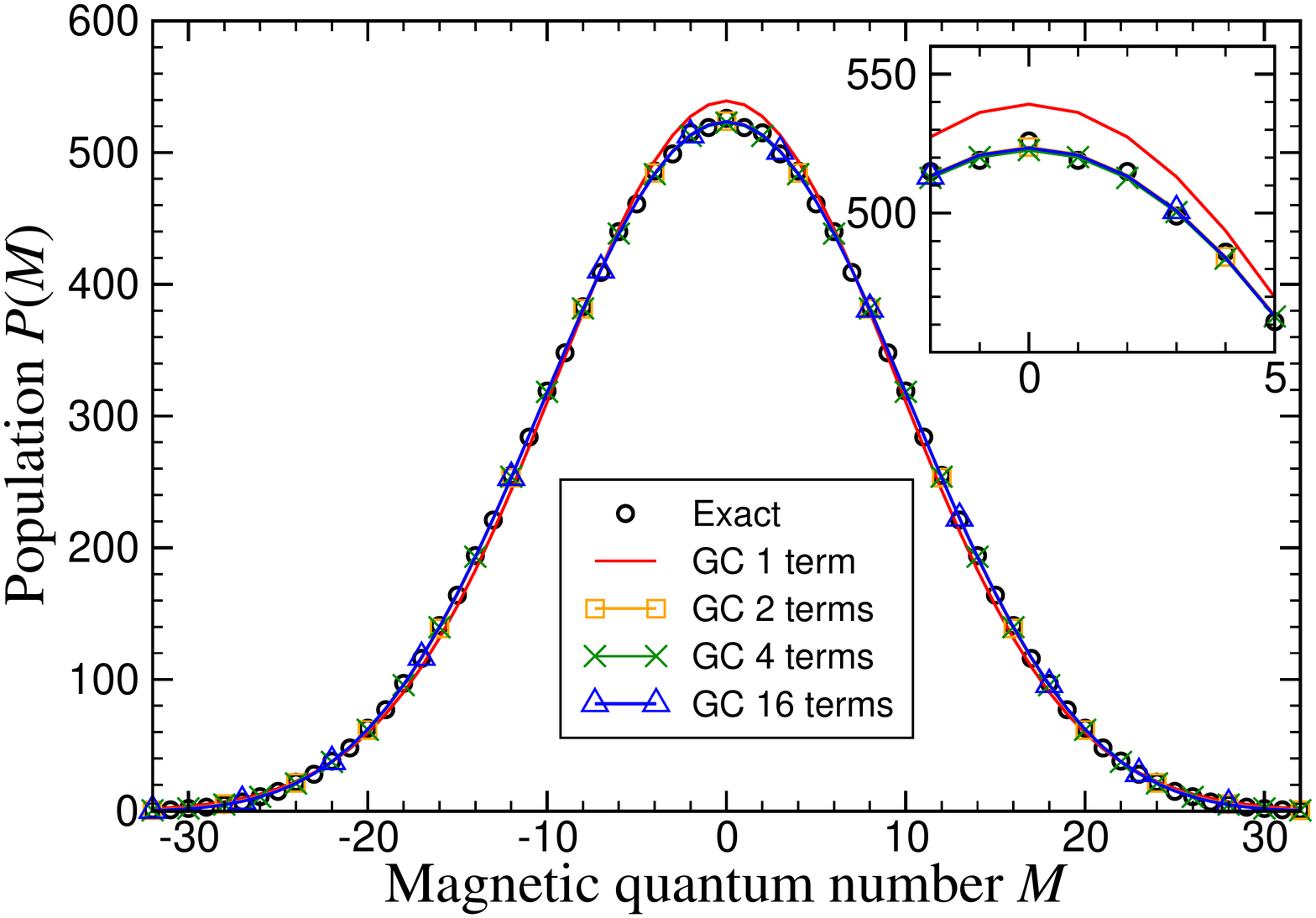}
}\goodgap%
\subfigure[Test of Gram-Charlier approximation as a function of the 
half-truncation order $k_\text{max}/2$. The green line with square 
symbols (resp. blue line with triangle symbols) is the absolute value  
$|P_\text{GC}(M;k_\text{max})-P(M)|$ for $M=\pm J_\text{max})$ (resp. 
$M=0$).]{\label{fig:error_GC_j15N8}
\includegraphics[scale=0.3]{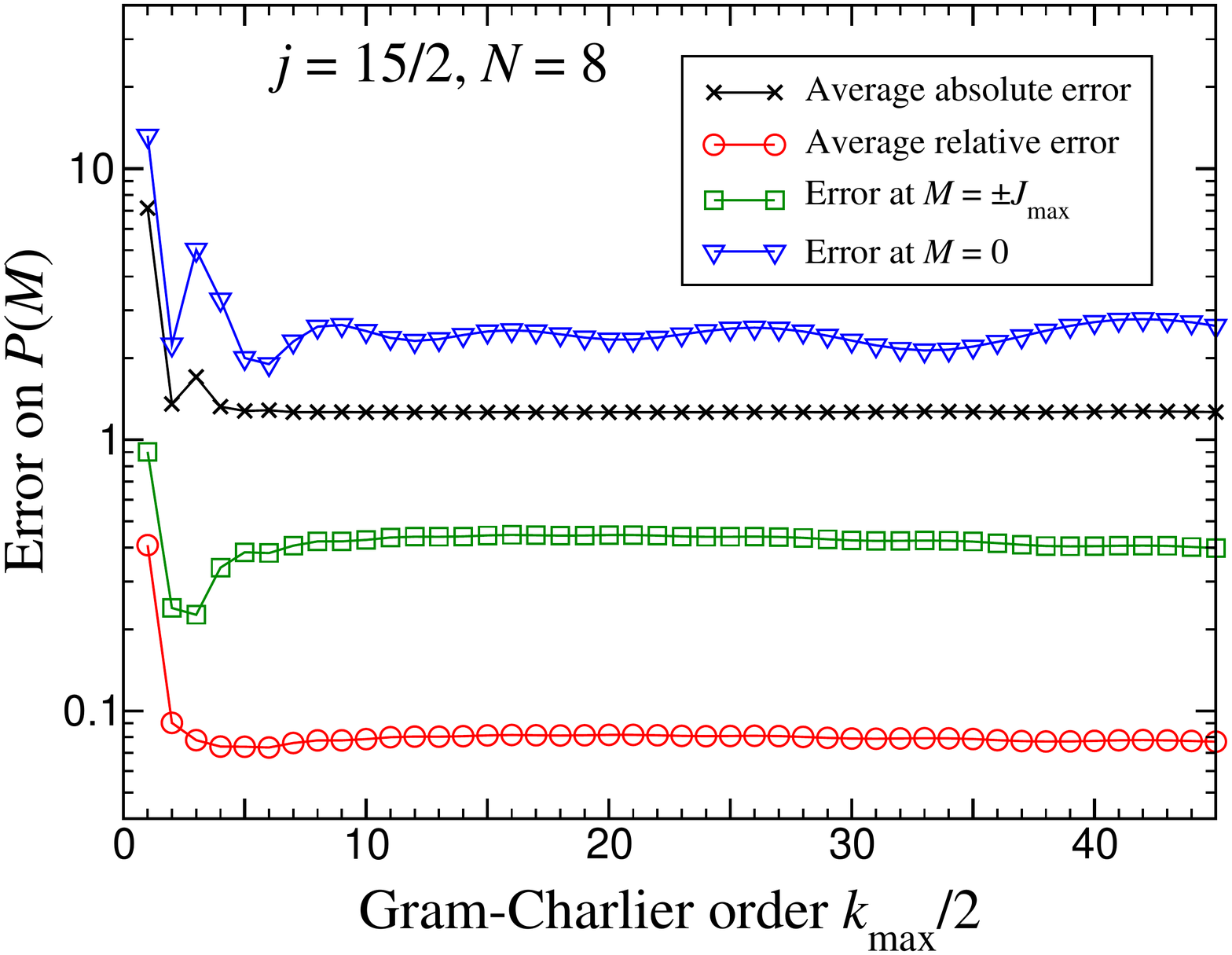}
}
\caption{Exact and Gram-Charlier approximation for the magnetic quantum 
number $P(M)$ in the relativistic configuration $j=15/2$, $N=8$.
\label{fig:GC_j15N8_distribM_err}}
\end{figure}

One may note that several works in plasma physics or EBIT spectroscopy 
deal with ions involving almost half-filled d or f subshells 
\cite{ZIGLER1987,RADTKE2001,Jonauskas2012}. Such subshells also deserve 
consideration in plasma sources for nanolithography \cite{SHEVELKO1998,%
OSullivan2015}.
We have plotted the exact and Gram-Charlier distributions for $P(M)$ in 
the half-filled subshell $N=j+1/2$ with $j=15/2$ on Fig. 
\ref{fig:PM_GC_j15N8}, for which $J_\text{max}=32$. 
One observes that the Gram-Charlier approximation performs well on the whole 
$M$-range. In more detail the simple 1-term form is accurate everywhere except 
close to the $M=0$ region, and the 2-term form, including variance and 
kurtosis, provides a fair approximation whatever $M$. In order to get a more 
quantitative picture, we have plotted in Fig. \ref{fig:error_GC_j15N8} the 
various evaluations of the error done as a function of the half truncation 
index $k/2$. On this figure the errors at $M=\pm J_\text{max}$ or $M=0$ 
are indeed the absolute differences $|P_\text{GC}(M;k_\text{max})-P(M)|$ 
to allow for a logarithmic scale, but it is noticeable that, for both $M$ 
values, the sign of the differences $P_\text{GC}(M;k_\text{max})-P(M)$ is 
positive for $k_\text{max}=2$, and negative for higher $k$. 
It turns out that including terms in the Gram-Charlier expansion 
beyond $k=4$ brings little improvement in the analytical representation of 
$P(M)$. It is even surprising that the various plotted errors tend to 
some asymptotic value, namely one notes that, for $M=J_\text{max}$, one has
$P_\text{GC}(M;k)-P(M)\rightarrow -0.4$ for large $k$, and for $M=0$ the 
``limit'' is $\sim -2.5$ with some oscillations. Therefore the ``convergence'' 
of the Gram-Charlier series is really poor in this case, the two-term 
expansion including up to the excess kurtosis providing a fair approximation 
for such half-filled subshells. This agrees with the conclusions obtained 
previously by Bauche \textit{et al} \cite{BAUCHE87}, though the effect of 
high-order terms was not quantitatively evaluated in this paper. Once again 
this case study suggests that the Gram-Charlier expansion provides an 
asymptotic representation of the magnetic momentum distribution.

\subsection{Example of multiple subshells}
\begin{figure}[htbp]
\centering
\includegraphics[scale=0.50,angle=0]{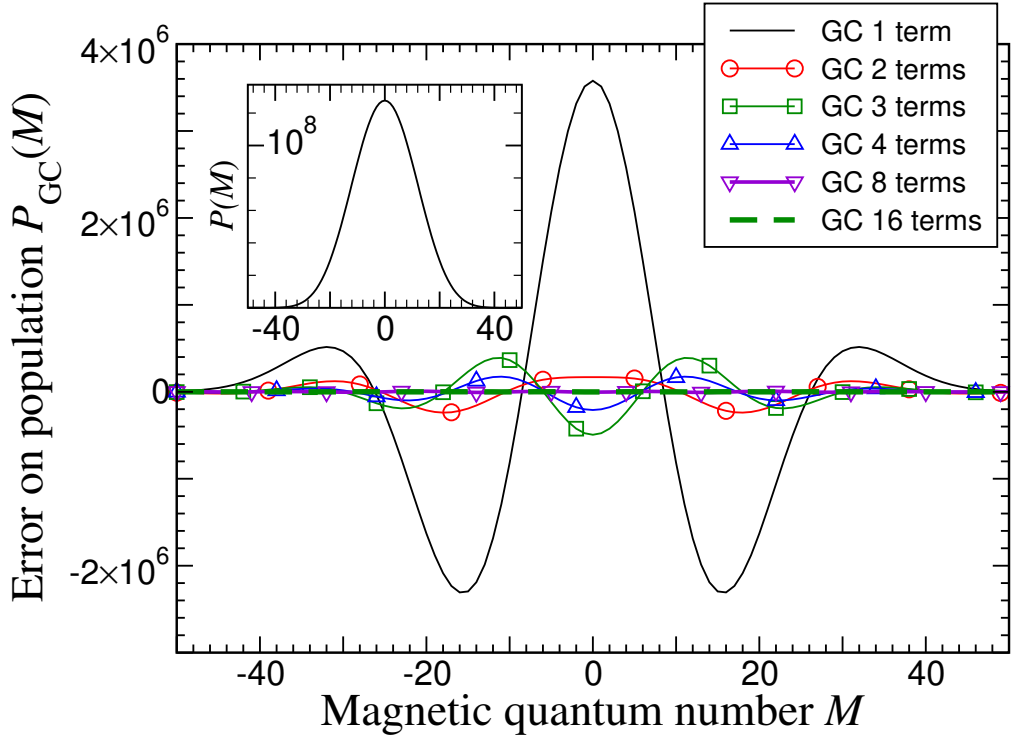}
\caption{Difference between Gram-Charlier approximation at various orders and 
exact value for the distribution $P(M)$. The relativistic configuration analyzed 
consists of 10 subshells $j=1/2$--$19/2$, all containing a single electron. 
The inset shows the exact distribution. \label{fig:PM_GC_j1j3-j19}}
\end{figure}

\begin{figure}[htbp]
\centering
\subfigure[Average error done in using the Gram-Charlier formula for the 
distribution $P(M)$ according to formulas (\ref{eq:abserr},\ref{eq:relerr})]
{\label{fig:error_GC_j1j3-j19}
\includegraphics[scale=0.3125]{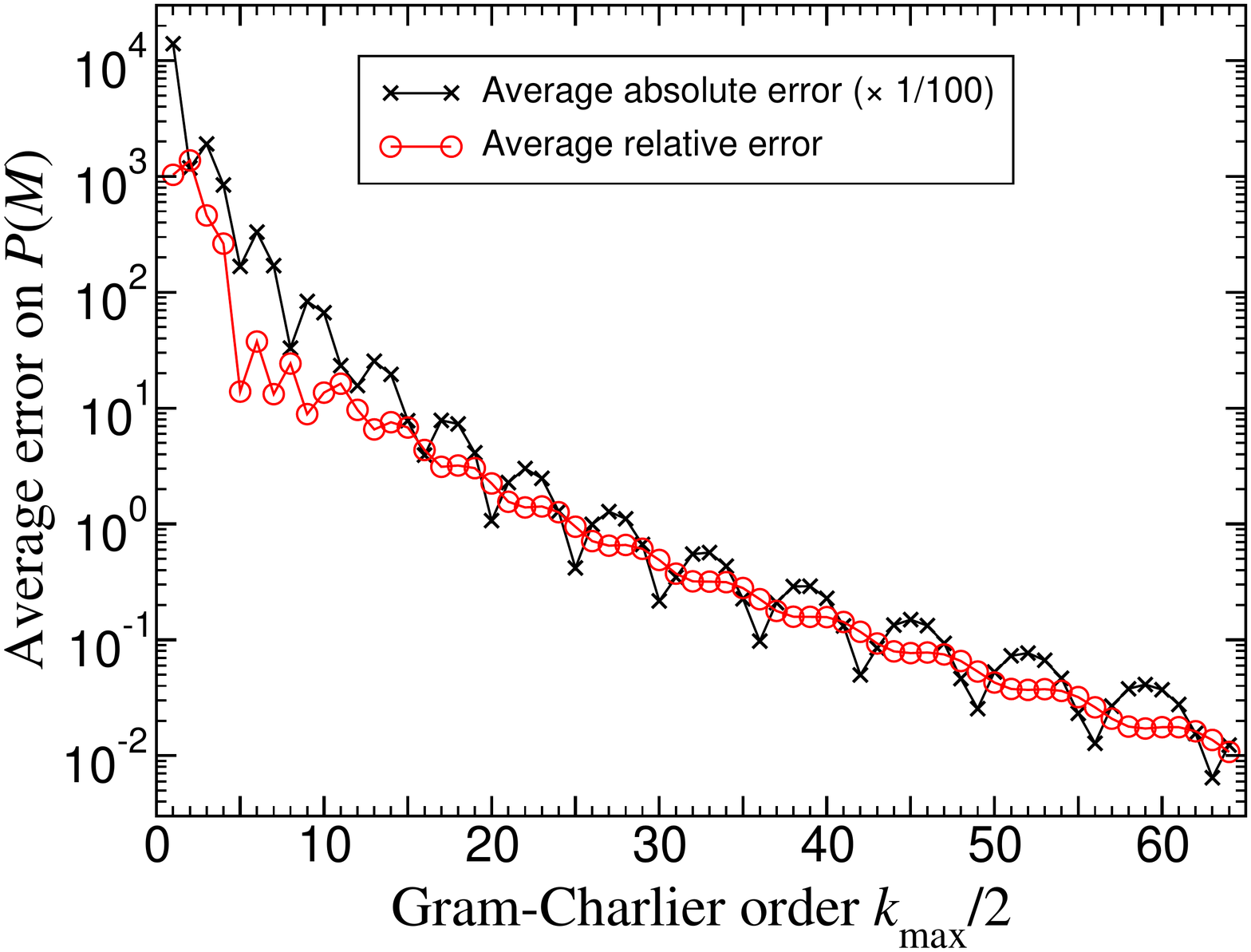}
}\goodgap%
\subfigure[Absolute error done in using the Gram-Charlier formula for the 
values $P(J_\text{max})$ and $P(0)$.]{\label{fig:errorM0Jmax_GC_j1j3-j19}%
\includegraphics[scale=0.3125]{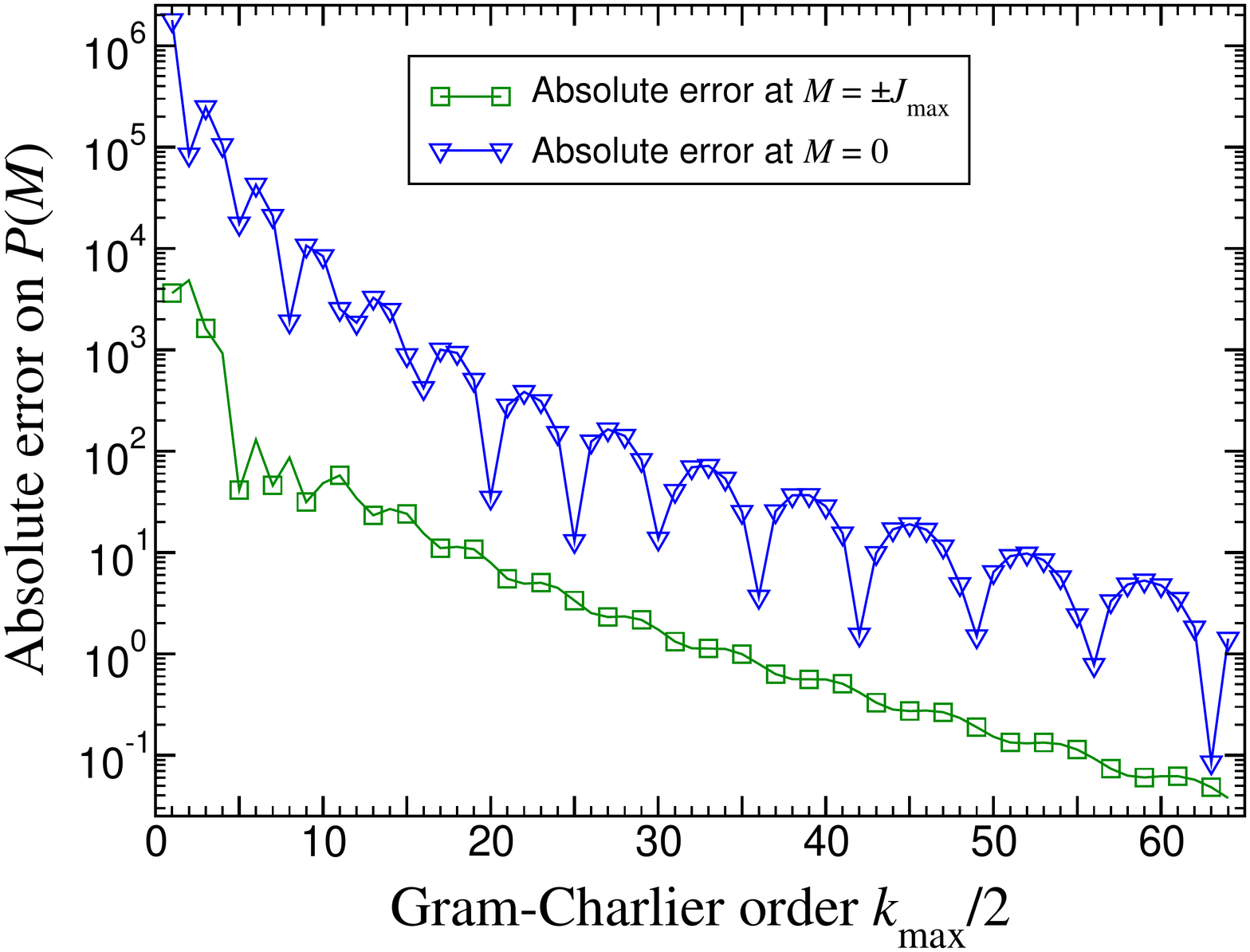}
}
\caption{Analysis of the Gram-Charlier series convergence for the 
$P(M)$ distribution in the relativistic configuration with 10 subshells 
$j=1/2$--$19/2$, all containing a single electron. 
\label{fig:GC_analysis_j1j3-j19}}
\end{figure}

It is not obvious to find situations where configurations with many open 
subshells contribute significantly to plasma spectra. However it is worth 
noting that the case of several singly-populated subshells is connected to 
the numbering of configurations contained in a superconfiguration analyzed 
in Ref.~\cite{Pain2020} because the cumulants are formally identical. 
Consequently, as a last illustration for the analysis of the 
$P(M)$ distribution we consider here a configuration 
with 10 subshells $j=1/2$--$19/2$, all containing a single electron. 
For this 10-electron configuration one has $J_\text{max}=50$, 
the degeneracy is $2^{10}.10!=3.7158912\times10^9$, and the population 
$P(M)$ varies on 8 orders of magnitude. We have plotted in 
Fig.~\ref{fig:PM_GC_j1j3-j19} the $P(M)$ distribution computed exactly and the 
differences between Gram-Charlier expansion truncated at various orders and 
the exact value. It turns out that the approximation with one term differs 
from the exact value, while Gram-Charlier approximation with at least 2 terms 
agrees with the exact value at the drawing accuracy. 
A more quantitative picture is provided by Figs \ref{fig:error_GC_j1j3-j19} 
and \ref{fig:errorM0Jmax_GC_j1j3-j19}. The former is a plot of the absolute 
and relative errors. The latter is the plot of the absolute difference 
between Gram-Charlier and exact values $\left|P_\text{GC}(M;k)-P(M)\right|$ 
for various $k$ values at $M=0$ and $M=J_\text{max}$. The differences 
$\left|P_\text{GC}(M;k)-P(M)\right|$ may also be divided by the exact $P(M)$ 
which are $P(0)=1.27707302\times10^8$ and $P(J_\text{max})=1$ respectively. 
Therefore from Fig. \ref{fig:errorM0Jmax_GC_j1j3-j19} it appears that the 
relative accuracy is much better for $M=0$ than for $M=J_\text{max}$.

As seen on these figures, the description of the distribution $P(M)$ by the 
Gram-Charlier expansion improves continuously with $k_\text{max}$. With 
16-byte floating point accuracy, we noticed a divergence on the absolute and 
relative errors for $k_\text{max}\simeq64$, while such behavior disappears 
in the present computations using Mathematica software. One notes that using 
$k_\text{max}=5$, the gain in accuracy versus an approximation including 
only the kurtosis ($k_\text{max}=5$) is significant, which gives a certain 
interest to the present analysis.


\subsection{Convergence of the Gram-Charlier series}
\begin{figure}[htb]
\centering
\subfigure[Configuration $j=7/2,N=2$]{\label{fig:cGCHe_j7n2}
\includegraphics[scale=0.325,angle=0]{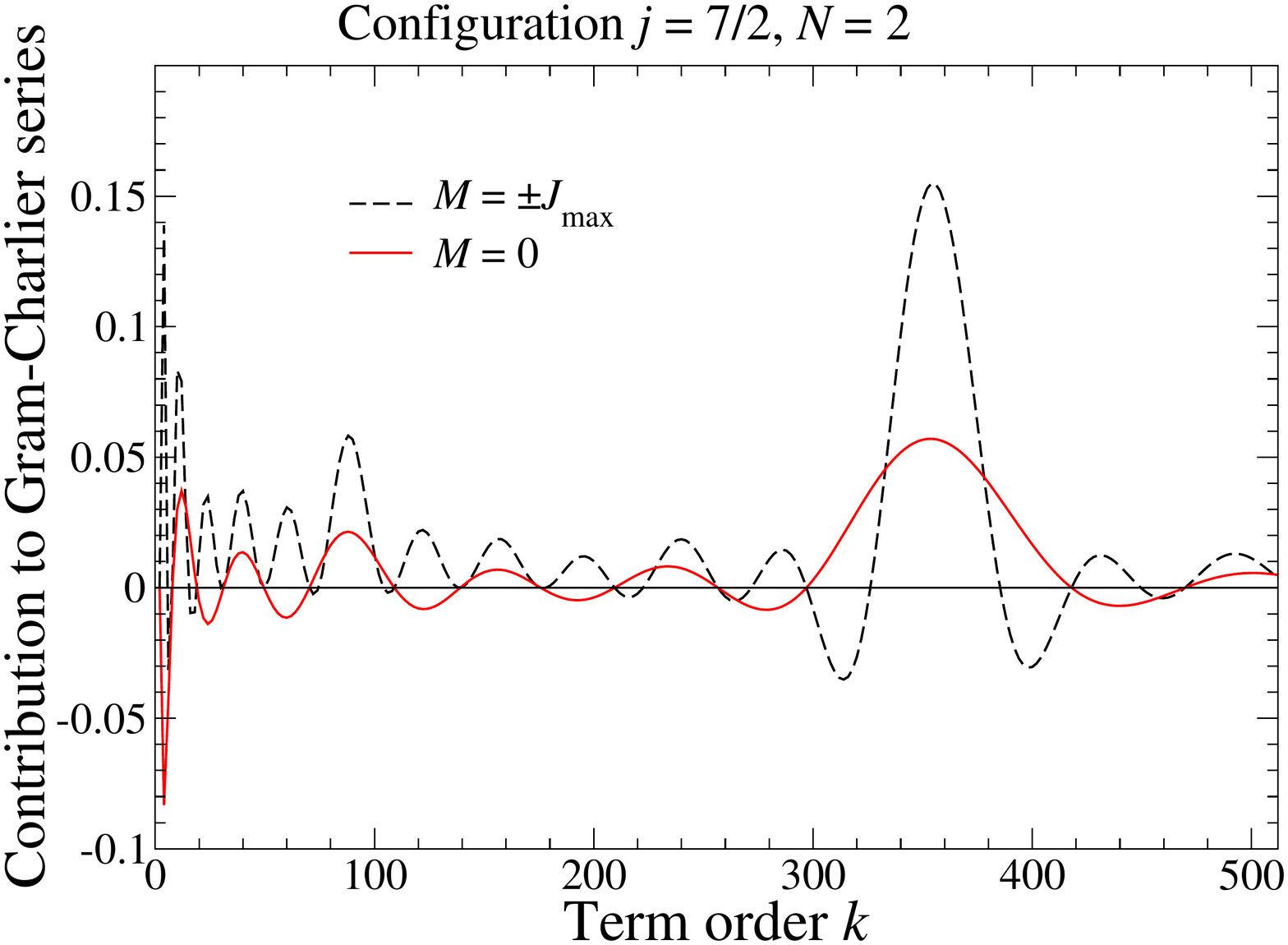}
}\\
\subfigure[Configuration $j=15/2,N=8$]{\label{fig:cGCHe_j15n8}
\includegraphics[scale=0.3125,angle=0]{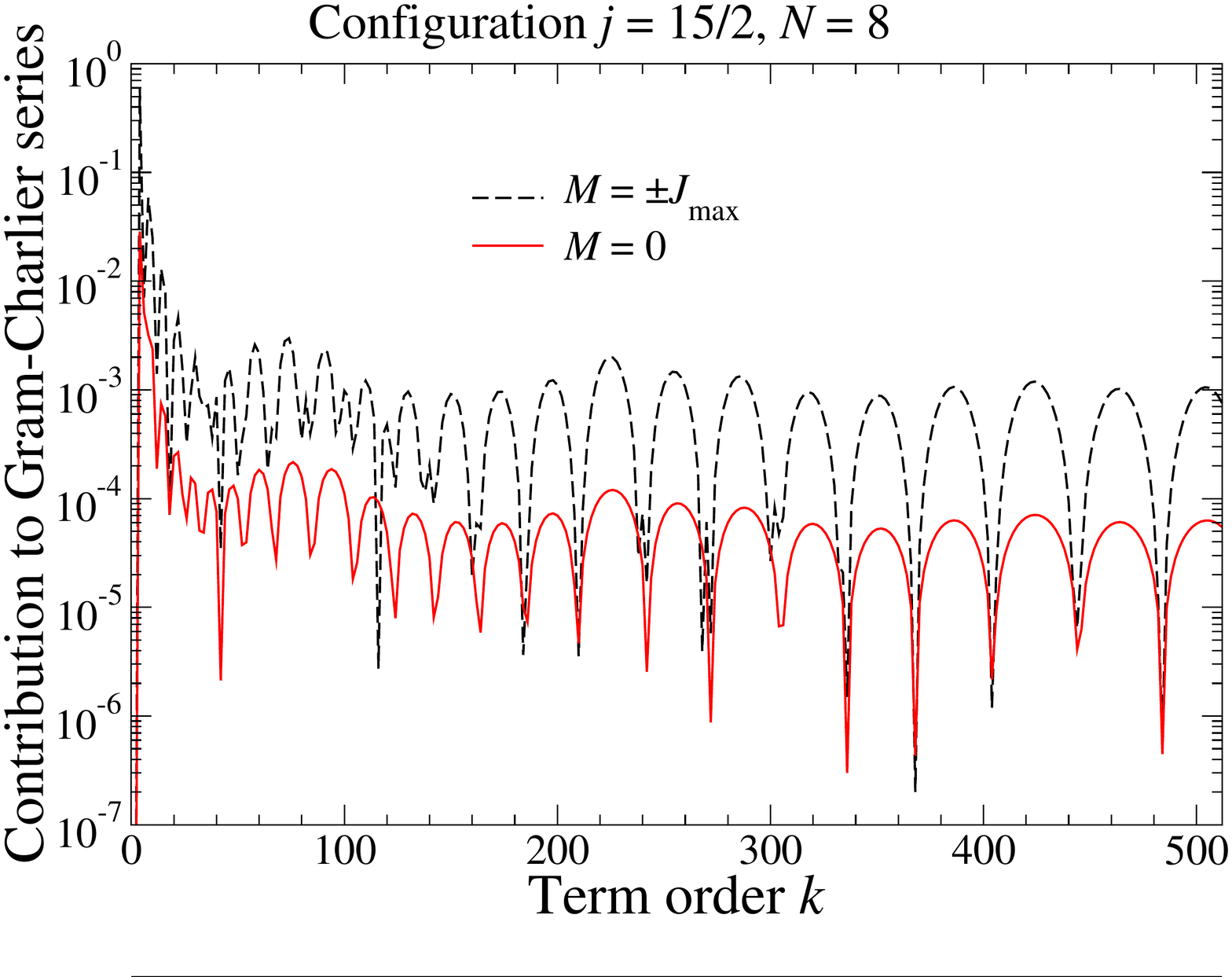}
}
\goodgap
\subfigure[Configuration with 5 subshells, $j_i=i-1/2, N_i=i$ ($i=$1--5)]{%
\label{fig:cGCHe_j1j3n2j5n3j7n4j9n5}
\includegraphics[scale=0.3125,angle=0]{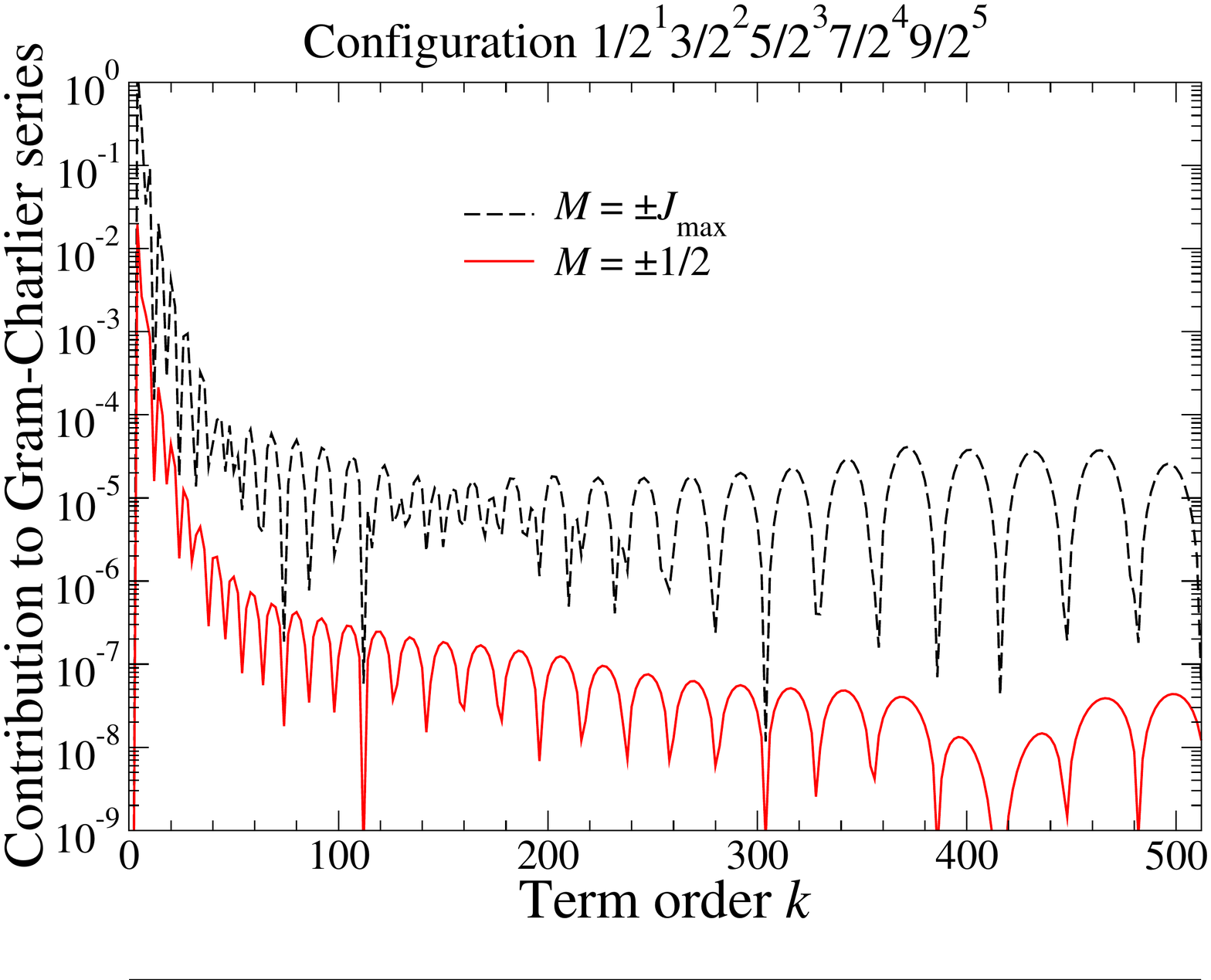}
}
\caption{Dependence of the generic term in the Gram-Charlier expansion 
$c_k He_k(M/\sigma)$ versus $k$ for three relativistic configurations 
and for two $M$ values. In the last two cases, the absolute value of this 
term is plotted in order to allow for logarithmic scale.\label{fig:cGCHe}}
\end{figure}
While we estimate the question of the mathematical convergence of the 
Gram-Charlier series to be outside the scope of this work, it is useful to 
check how the generic term of the sum in Eq. (\ref{eq:Gram-Charlier}) varies 
with $k$. To this respect, we have plotted in Fig.~\ref{fig:cGCHe} 
the term $c_k He_k(M/\sigma)$ or its absolute value versus $k$ for the 
values $M=J_\text{max}$ and $M=0$ or $M=\pm1/2$ for three configurations. 
Of course these computations were performed with arbitrary precision 
software to avoid inaccuracies when computing large-order coefficients. 
In Fig.~\ref{fig:cGCHe_j7n2} illustrating a 2-electron configuration case we 
notice that the $c_kHe_k$ term oscillates with $k$ and do not decrease in 
absolute value below 0.1. For the more populated configurations shown in 
Fig.~\ref{fig:cGCHe_j15n8} (resp. \ref{fig:cGCHe_j1j3n2j5n3j7n4j9n5}) the 
generic term of the series also oscillates and decreases to lower values. 
One notices a plateau in the oscillation amplitudes at $10^{-4}$ for $M=0$ 
(resp. $10^{-7}$ for $M=1/2$). However, as far as we could check, we did not 
observe a subsequent decrease of this generic term for greater $k$ values. 
These numerical considerations lead us to estimate that the Gram-Charlier 
series is probably not convergent, though accounting for a large number of 
terms may significantly improve the quality of this approximation, with 
better results for configurations with a large number of electrons. This 
behavior is characteristic of an asymptotic expansion. 

\section{Edgeworth series}\label{sec:Edgeworth}
\subsection{Definition}
As mentioned by various authors \cite{Blinnikov1998,deKock2011} some 
statistical distributions are better represented by Edgeworth series than by 
Gram-Charlier series. The Edgeworth distribution of the variable $X$ is 
naturally expressed in terms of cumulants, and is written as an expansion 
versus powers of the standard deviation
\begin{subequations}\begin{equation}
 E(X) = G\frac{\exp(-x^2/2)}{\sqrt{2\pi}\sigma} 
  \left[ 1+\sum_{s=1}^\infty\sigma^s\sum_{\{k_m\}}He_{s+2r}(x)
  \prod_{m=1}^s\frac{1}{k_m!}\left(\frac{S_{m+2}}{(m+2)!}\right)^{k_m} 
  \right]\label{eq:Edgeworth}
\end{equation}
where we have introduced the reduced variable $x=(X-\left<X\right>)/\sigma$ 
and the modified cumulants $S_n=\kappa_n/\sigma^{2n-2}$. The set of indices 
$\{k_m\}$ refer to all $s$-tuples verifying 
\begin{gather} 
r=k_1+k_2+\cdots k_s\\
k_1+2k_2+\cdots+sk_s=s,
\end{gather}\end{subequations}
i.e., partitions of the integer $s$. Since the analyzed distribution $P(M)$ 
is even, this series involves only even $s$ orders. An inspection of the 
above formulas shows that the $s=1$ contribution in the sum is proportional 
to the asymmetry $\kappa_3$ which cancels for the $P(M)$ distribution. 
The $s=2$ term is proportional to the excess kurtosis $\kappa_4$ and is 
identical to the first correction in the Gram-Charlier series. More 
generally one can check that the sum of coefficients factoring a polynomial 
of given order $He_{k}(x)$ in the expansion (\ref{eq:Edgeworth}) is equal 
to the coefficient of the same order $c_{k}$ in the Gram-Charlier
series. This property has been used to check the consistency of the 
coefficients in these expansions. It means that the Edgeworth series is 
indeed a rearrangement of the Gram-Charlier series, where each individual 
term is recast in various orders of the Edgeworth series. 

\subsection{A test of Edgeworth accuracy}
\begin{figure}[htbp]
\centering
\includegraphics[scale=0.50,angle=0]%
{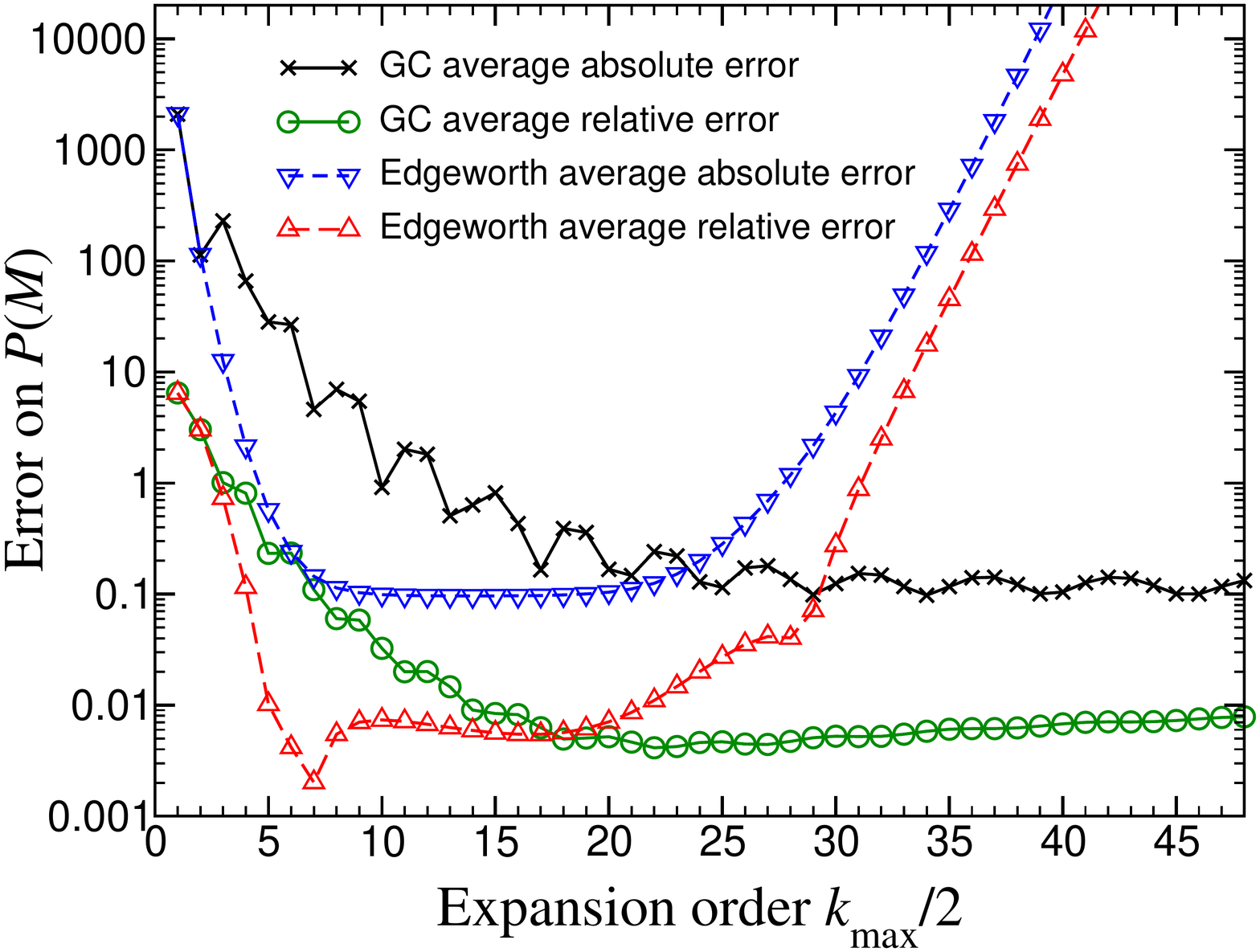}
\caption{Absolute error and relative done in using the Gram-Charlier 
and Edgeworth formula for the $P(M)$ distribution. The analyzed configuration 
is $j_1=1/2, N_1=1, j_2=3/2, N_2=2, j_3=5/2, N_3=3, j_4=7/2, N_4=4, j_5=9/2, 
N_5=5$.\label{fig:GC_Edgeworth_j1j3j5j7j9}}
\end{figure}

The relative efficiency of Edgeworth and Gram-Charlier series is illustrated 
in Fig. \ref{fig:GC_Edgeworth_j1j3j5j7j9} for the configuration with 5 
half-filled subshells $j_1=1/2, N_1=1, j_2=3/2, N_2=2, j_3=5/2, N_3=3, 
j_4=7/2, N_4=4, j_5=9/2, N_5=5$, where we have plotted the errors 
(\ref{eq:abserr}) and (\ref{eq:relerr}) for both series as functions of the 
half truncation index $k_\text{max}/2$. For the lowest $k_\text{max}$ values, 
the Edgeworth series provides a better approximation by a factor of 10. 
For greater values of this index, the Gram-Charlier expansion tends toward an 
acceptable approximation, with a relative error below 0.01. Conversely, the 
Edgeworth expansion is clearly divergent for $k_\text{max}\gtrsim 40$ or 50. 

This expansion has been studied numerically using high-precision 
arithmetic in Fortran and arbitrary precision in Mathematica, which allows us 
to conclude that the divergence is not an artifact due to loss of accuracy in 
numerical computations but really a mathematical divergence. This behavior is 
similar to that observed for Gram-Charlier expansion, with two noticeable 
differences: the ``best-convergence plateau'' is reached earlier in the present 
case, and the onset of the divergence is also earlier and more pronounced here.

\section{Discussion and conclusion}\label{sec:conclu}
We have studied in this paper various aspects of the statistics of the 
total quantum magnetic number distribution $P(M)$ in the most general 
relativistic configuration, accounting for the fermion character of the 
electrons. We have mentioned that atomic configurations considered  
here may be of importance in several domains such as plasma spectroscopy, 
nanolithography-source design or EBIT facilities.
Up to our knowledge, there exists no compact analytical expression for the 
quantum magnetic number distribution, which justifies the present effort. 
Using the cumulant generating function formalism we have derived a 
recurrence relation for this function connecting four adjacent $j$ and $N$ 
values in the case of a single subshell $j^N$. This relation allowed us to 
establish a compact analytical expression for the cumulant generating 
function, which is straightforwardly generalized to a relativistic 
configuration with any subshell number. In the case of a single subshell this 
generating function allows us to express $P(M)$ as a n-th derivative. This 
formal property leads to two recurrence relations on $P(M)$ for adjacent $N$ 
or $j$, $j$ being allowed to span integer as well as half-integer values. 
Such recurrences prove to be quite efficient in obtaining the whole $P(M)$ 
distribution for complex configurations. We have been able to express the 
cumulants of the $P(M)$ distribution at any order for the most general 
relativistic configuration. This allowed us to build a Gram-Charlier 
approximation of the magnetic quantum number distribution at any order. 
The Gram-Charlier analysis performed here has provided a variety of results. 
First, it has been stressed that the handling of series with several tens 
of terms requires the use of arbitrary precision, since a ``divergence'' due 
to a loss of numerical accuracy may be observed for a $k_\text{max}\sim 40$.
For a subshell with significant population --- e.g., an half-filled subshell 
with large $j$ --- the first two terms of the expansion provide a very good 
approximation while adding many more terms do not improve the approximation 
at all. Conversely, in the cases with a large number of subshells each one 
with a small population, the quality of the Gram-Charlier expansion improves 
as more terms are added. Similar conclusions holds for ``exotic'' 
configurations for which the $M$-distribution shows a broad plateau, which 
can be fairly reproduced by including several tens of terms.

It has been verified that configurations with a large number of electrons 
are better represented by Edgeworth expansion for a moderate value of the 
truncation index, reaching a best-approximation plateau before the 
Gram-Charlier expansion. Furthermore, a better accuracy is achieved for 
configurations with a large number of electrons. However both expansions 
appear to be asymptotic and not convergent, with an earlier divergence for 
the Edgeworth series. Such conclusions are similar to what we obtained when 
considering the statistics of configurations inside a superconfiguration. 
A physically important application of this analysis is that it leads to 
useful information on the distribution of total angular momentum $J$ and on 
line numbers. This will be considered in a forthcoming paper.

\appendix
\section{Alternate proposal for a derivation of the expression of the 
cumulant generating function}\label{sec:relsNsN-1}
In order to prove the general expression (\ref{eq:exps}) one may directly 
establish the two-term relation (\ref{eq:relSNsN-1}) with
\begin{equation}
s_N(t) = \sum_M P(M)e^{Mt}=\sum_{m_1<m_2<\cdots<m_N}e^{(m_1+\cdots+m_N)t}.
\end{equation}
The recurrence (\ref{eq:relSNsN-1}) may be written
\begin{equation}
 \left(e^{(j+1-N/2)t}-e^{-(j+1-N/2)t}\right)s_{N-1}(t) 
  = \left(e^{Nt/2}-e^{-Nt/2}\right)s_{N}(t)
\end{equation}
or after some elementary transformation
\begin{equation}
 \left(e^{-(j+1-N)t}+e^{-(j-N)t}+\cdots+e^{jt}\right)s_{N-1}(t)
  = \left(1+e^t+\cdots+e^{(N-1)t}\right)s_{N}(t).
\end{equation}
The left member contains $(2j+2-N)\binom{2j+1}{N-1}$ terms and the right 
member contains $N\binom{2j+1}{N}$ terms which are both equal. A detailed 
inspection of the terms in both members shows that they are indeed identical. 
However this verification is somewhat tedious and the proof given in Section 
\ref{sec:mag} 
is easier to establish.

\section{Derivation of recurrence relations using generalized Pascal-triangle 
relations for the Gaussian binomial coefficients}\label{sec:rec_Gaussbin}
From the definition of the Gaussian binomial coefficients (\ref{eq:FjNz}), after 
elementary algebraic operations, we get the well-known triangle-like relation
\begin{equation}
\qbin{2j+1}{N}{z}=z^N\qbin{2j}{N}{z}+\qbin{2j}{N-1}{z}
\end{equation}
and, deriving this equation $n$ times versus $z$ using the Leibniz rule, we obtain
\begin{equation}
\left.\frac{\partial^n}{\partial z^n}\qbin{2j+1}{N}{z}\right|_{z=0}=
 \left.\sum_{k=0}^N\binom{n}{k}\frac{\partial^{n-k}}{\partial z^{n-k}}z^N.
 \frac{\partial^k}{\partial z^k}\qbin{2j}{N}{z}\right|_{z=0}
 +\left.\frac{\partial^n}{\partial z^n}\qbin{2j}{N-1}{z}\right|_{z=0}.
\end{equation}
From the value of the n-th derivative of $z^N$ at the origin, and after 
dividing by $n!$, we get
\begin{equation}
\frac{1}{n!}\left.\frac{\partial^n}{\partial z^n}\qbin{2j+1}{N}{z}\right|_{z=0}=
\left.\frac{1}{(n-N)!}.
\left. \frac{\partial^{n-N}}{\partial z^{n-N}}\qbin{2j}{N}{z}\right|_{z=0}
 +\frac{1}{n!}\frac{\partial^n}{\partial z^n}\qbin{2j}{N-1}{z}\right|_{z=0},
\end{equation}
so that, with the definition (\ref{eq:PM_derivn}) of $P(M)$ as a multiple 
derivative and the substitution $M=n-N(2j+1-N)/2$,
\begin{equation}\label{eq:recPM_tri1}
P(M;j,N)=P\left(M-\frac{N}{2};j-\frac{1}{2},N\right)+
 P\left(M+\frac{2j+1-N}{2};j-\frac{1}{2},N-1\right).
\end{equation}
With $0\leq n\leq 2J_{\mathrm{max}}$, using the notation $\mathscr{P}_{j,N}(n)$ 
introduced in Eq. (\ref{eq:Pscr_def}), we obtain
\begin{equation}\label{eq:recPn_tri1}
\mathscr{P}_{j,N}\left(n\right)=
 \mathscr{P}_{j-1/2,N}\left(n-N\right)+\mathscr{P}_{j-1/2,N-1}\left(n\right).
\end{equation}
It is also quite simple to derive from the definition (\ref{eq:FjNz}) another 
triangle-like equation
\begin{equation}
\qbin{2j+1}{N}{z}=\qbin{2j}{N}{z}+z^{2j+1-N}\qbin{2j}{N-1}{z}
\end{equation}
and therefore, applying the same procedure as above, we get
\begin{equation}
\left.\frac{\partial^n}{\partial z^n}\qbin{2j+1}{N}{z}\right|_{z=0}=
 \left.\frac{\partial^n}{\partial z^n}\qbin{2j}{N}{z}\right|_{z=0}
  +\left.\binom{n}{2j+1-N}\frac{\partial^{2j+1-N}}{\partial z^{2j+1-N}}z^{2j+1-N}.
  \frac{\partial^{n-2j-1+N}}{\partial z^{n-2j-1+N}}\qbin{2j}{N-1}{z}\right|_{z=0},
\end{equation}
i.e.
\begin{equation}\label{eq:recPM_tri2}
P(M;j,N)=P\left(M+\frac{N}{2};j-\frac{1}{2},N\right)+
 P\left(M-\frac{2j+1-N}{2};j-\frac{1}{2},N-1\right).
\end{equation}
Since the distribution $P(M)$ is even, one easily checks that this recurrence 
relation is indeed equivalent to the previous one (\ref{eq:recPM_tri1}).
With $0\leq n\leq 2J_{\mathrm{max}}$ and using the $\mathscr{P}_{j,N}$ 
notation, we obtain 
\begin{equation}\label{eq:recPn_tri2}
\mathscr{P}_{j,N}\left(n\right)=
 \mathscr{P}_{j-1/2,N}\left(n\right)+\mathscr{P}_{j-1/2,N-1}\left(n-2j-1+N\right).
\end{equation}
Combining Eqs. (\ref{eq:recPn_tri1}) and (\ref{eq:recPn_tri2}), one gets 
\begin{equation}
\mathscr{P}_{j-1/2,N}\left(n-N\right)+\mathscr{P}_{j-1/2,N-1}\left(n\right)=
 \mathscr{P}_{j-1/2,N}\left(n\right)+\mathscr{P}_{j-1/2,N-1}\left(n-2j-1+N\right),
\end{equation}
which is exactly the recurrence relation over $N$ (\ref{eq:recOPM_overN}) 
with the substitution $j\rightarrow j-1/2$. Similarly, making the substitution 
$n\rightarrow n-2j-1+N$ in Eq. (\ref{eq:recPn_tri1}), we get
\begin{equation}\label{eq:recPn_tri1b}
\mathscr{P}_{j-1/2,N-1}\left(n-2j-1+N\right)=\mathscr{P}_{j,N}\left(n-2j-1+N\right)
 -\mathscr{P}_{j-1/2,N}\left(n-2j-1\right).
\end{equation}
Combining Eq. (\ref{eq:recPn_tri2}) with Eq. (\ref{eq:recPn_tri1b}) yields
exactly the recurrence relation over $j$ (\ref{eq:recOPM_overj}).

\section{Explicit values for the moments in a relativistic configuration}
\label{sec:appmomts}
The explicit value for the cumulants (\ref{eq:cumul}) and the relation 
between moments and cumulants (\ref{eq:momts_cumults}) shows that the 
polynomial form of the moments inside a $j^N$-subshell is written as
\begin{equation}
 \mu_{2k}=\sum_{p=1}^{k}M(2k,p)\left[N(2j+1-N)\right]^p.
\end{equation}
The first values for $M(2k,p)$ are listed below. 
\begingroup
\allowdisplaybreaks
\begin{subequations}\begin{align}
M(2,1)&=\frac{j+1}{6}\\
M(4,1)&=-\frac{1}{30}(j+1)^2(2j+1)\\
M(4,2)&=\frac{1}{60}(j+1)(5j+6)\\
M(6,1)&=\frac{1}{126}(j+1)^2(2j+1)\left(8j^2+12j+3\right)\\
M(6,2)&=-\frac{1}{252}(j+1)(2j+1)\left(21j^2+50j+30\right)\\
M(6,3)&=\frac{1}{504}(j+1)\left(35j^2+91j+60\right)\\
M(8,1)&=-\frac{1}{90}(j+1)^2(2j+1)\left(24j^4+72j^3+70j^2+24j+3\right)\\
M(8,2)&=\frac{1}{540}(j+1)(2j+1)\left(202j^4+815j^3+1177j^2+693j+126\right)\\
M(8,3)&= -\frac{1}{540}(j+1)(2j+1)(3j+4)\left(35j^2+92j+63\right)\\
M(8,4)&= \frac{1}{2160}(j+1)(5j+7)\left(35j^2+98j+72\right)\\
M(10,1)&=\frac{1}{66}(j+1)^2(2j+1)\left(4j^2+6j+1\right)
 \left(32j^4+96j^3+92j^2+30j+5\right)\\
M(10,2)&=-\frac{1}{396}(j+1)(2j+1)
 \left(1144j^6+6488j^5+14634j^4+16450j^3+9419j^2+2520j+270\right)\\
M(10,3)&= \frac{1}{792}(j+1)(2j+1)
 \left(1342j^5+7755j^4+17546j^3+19076j^2+9599j+1620\right)\\
M(10,4)&= -\frac{1}{792}(j+1)(2j+1)\left(385j^4+2211j^3+4840j^2+4784j+1800\right)\\
M(10,5)&= \frac{1}{3168}(j+1)\left(385j^4+2310j^3+5291j^2+5478j+2160\right)\\
M(12,1)&=-\frac{691}{8190}(j+1)^2(2j+1) \nonumber\\
 &\quad\times\left(256j^8+1536j^7+3712j^6+4608j^5+3160j^4+1272j^3+338j^2+48j+3\right)\\
M(12,2)&=\frac{1}{49140}(j+1)(2j+1)\nonumber\\
 &\quad\times\left(1663792j^8+12141368j^7+37081404j^6+60997498j^5+58017896j^4 \right.\nonumber\\
 &\qquad \left. +32097489j^3+10156029j^2+1829949j+136818\right)\\
M(12,3) &= -\frac{1}{98280}(j+1)(2j+1)\nonumber\\
 &\quad\times\left(2126124j^7+16084328j^6+50719883j^5+85218445j^4+80675641j^3 \right.\nonumber\\
 &\qquad \left.+41715213j^2+10549422j+1094544\right) \\
M(12,4) &= \frac{1}{196560}(j+1)(2j+1)\left(1431430j^6+11062051j^5+35328306j^4 \right.\nonumber\\
 &\quad \left.+59068682j^3+53560409j^2+24090798j+3830904\right) \\
M(12,5) &= -\frac{1}{393120}(j+1)(2j+1)\nonumber\\
 &\quad\times\left(525525j^5+4059055j^4+12765753j^3+20419373j^2+16593534j+5472720\right) \\
M(12,6) &= \frac{1}{786240}(j+1)\nonumber\\
 &\quad\times\left(175175j^5+1401400j^4+4569565j^3+7583576j^2+6396156j+2189088\right)
\end{align}\end{subequations}
\endgroup
One notes that as a rule the moments exhibit a more complex expression than the 
cumulants.

\section{A recurrence relation on the distribution moments}\label{sec:rec_mom}
From the expressions of the moments (\ref{eq:defmt}) and of the generating 
function (\ref{eq:defs}) one gets the non-normalized moments 
\begin{equation}
\mathscr{M}_k(j^N) = \binom{2j+1}{N}\mu_{k}
 =\left.\frac{\partial^k}{\partial t^k}s(N,j,t)\right|_{t=0}.
\end{equation} 
After deriving Eq. (\ref{eq:devSN}) $k$ times versus $t$ using the 
Leibniz rule, one gets the relation
\begin{equation}
\mathscr{M}_k(j^N)=\frac{1}{N}\sum_{p=1}^{N}(-1)^{p-1}
\sum_{q=0}^k\binom{k}{q}p^{k-q}\mathscr{M}_q(j^{N-p})\mathscr{M}_{k-q}(j^1)
\end{equation}
and for the normalized moments 
\begin{equation}
\mu_k(j^N)=\frac{2j+1}{N}\sum_{p=1}^{N}(-1)^{p-1}
\frac{N!(2j+1-N)!}{(N-p)!(2j+1-N+p)!}
\sum_{q=0}^k\binom{k}{q}p^{k-q}\mu_q(j^{N-p})\mu_{k-q}(j^1)
\end{equation}
which indeed involves only even indices $k,q$. In the above equations one 
must define the moments for the empty subshell $j^0$, which are 
$\mathscr{M}_s(j^0)=\mu_s(j^0)=0$ if $s>0$, and $\mathscr{M}_0(j^0)=
\mu_0(j^0)=1$.
The even-order moments for a single-electron subshell $j^1$, defined as
$\mathscr{M}_{2k}(j^1)=\sum_{m=-j}^j m^{2k}$ are simply related to Bernoulli 
numbers $B_n(x)$ 
\begin{equation}
\mathscr{M}_{2k}(j^1)=\frac{B_{2k+1}(j+1)-B_{2k}(-j)}{2k+1}.
\end{equation}

\bibliography{jrnlabbr,M-distrib}

\begin{thebibliography}{39}%
\makeatletter
\providecommand \@ifxundefined [1]{%
 \@ifx{#1\undefined}
}%
\providecommand \@ifnum [1]{%
 \ifnum #1\expandafter \@firstoftwo
 \else \expandafter \@secondoftwo
 \fi
}%
\providecommand \@ifx [1]{%
 \ifx #1\expandafter \@firstoftwo
 \else \expandafter \@secondoftwo
 \fi
}%
\providecommand \natexlab [1]{#1}%
\providecommand \enquote  [1]{``#1''}%
\providecommand \bibnamefont  [1]{#1}%
\providecommand \bibfnamefont [1]{#1}%
\providecommand \citenamefont [1]{#1}%
\providecommand \href@noop [0]{\@secondoftwo}%
\providecommand \href [0]{\begingroup \@sanitize@url \@href}%
\providecommand \@href[1]{\@@startlink{#1}\@@href}%
\providecommand \@@href[1]{\endgroup#1\@@endlink}%
\providecommand \@sanitize@url [0]{\catcode `\\12\catcode `\$12\catcode
  `\&12\catcode `\#12\catcode `\^12\catcode `\_12\catcode `\%12\relax}%
\providecommand \@@startlink[1]{}%
\providecommand \@@endlink[0]{}%
\providecommand \url  [0]{\begingroup\@sanitize@url \@url }%
\providecommand \@url [1]{\endgroup\@href {#1}{\urlprefix }}%
\providecommand \urlprefix  [0]{URL }%
\providecommand \Eprint [0]{\href }%
\providecommand \doibase [0]{https://doi.org/}%
\providecommand \selectlanguage [0]{\@gobble}%
\providecommand \bibinfo  [0]{\@secondoftwo}%
\providecommand \bibfield  [0]{\@secondoftwo}%
\providecommand \translation [1]{[#1]}%
\providecommand \BibitemOpen [0]{}%
\providecommand \bibitemStop [0]{}%
\providecommand \bibitemNoStop [0]{.\EOS\space}%
\providecommand \EOS [0]{\spacefactor3000\relax}%
\providecommand \BibitemShut  [1]{\csname bibitem#1\endcsname}%
\let\auto@bib@innerbib\@empty
\bibitem [{\citenamefont {Bethe}(1936)}]{BETHE36}%
  \BibitemOpen
  \bibfield  {author} {\bibinfo {author} {\bibfnamefont {H.~A.}\ \bibnamefont
  {Bethe}},\ }\bibfield  {title} {\bibinfo {title} {An attempt to calculate the
  number of energy levels of a heavy nucleus},\ }\href
  {https://doi.org/10.1103/PhysRev.50.332} {\bibfield  {journal} {\bibinfo
  {journal} {Phys. Rev.}\ }\textbf {\bibinfo {volume} {50}},\ \bibinfo {pages}
  {332} (\bibinfo {year} {1936})}\BibitemShut {NoStop}%
\bibitem [{\citenamefont {Cowan}(1981)}]{COWAN81}%
  \BibitemOpen
  \bibfield  {author} {\bibinfo {author} {\bibfnamefont {R.~D.}\ \bibnamefont
  {Cowan}},\ }\href@noop {} {\emph {\bibinfo {title} {The Theory of Atomic
  Structure and Spectra}}},\ Los Alamos Series in Basic and Applied Sciences\
  (\bibinfo  {publisher} {University of California Press, Ltd.},\ \bibinfo
  {address} {Berkeley},\ \bibinfo {year} {1981})\BibitemShut {NoStop}%
\bibitem [{\citenamefont {Judd}(1968)}]{JUDD68}%
  \BibitemOpen
  \bibfield  {author} {\bibinfo {author} {\bibfnamefont {B.~R.}\ \bibnamefont
  {Judd}},\ }\bibfield  {title} {\bibinfo {title} {Atomic term patterns},\
  }\href {https://doi.org/10.1103/PhysRev.173.39} {\bibfield  {journal}
  {\bibinfo  {journal} {Phys. Rev.}\ }\textbf {\bibinfo {volume} {173}},\
  \bibinfo {pages} {39} (\bibinfo {year} {1968})}\BibitemShut {NoStop}%
\bibitem [{\citenamefont {Breit}(1926)}]{BREIT26}%
  \BibitemOpen
  \bibfield  {author} {\bibinfo {author} {\bibfnamefont {G.}~\bibnamefont
  {Breit}},\ }\bibfield  {title} {\bibinfo {title} {An application of {P}auli's
  method of coordination to atoms having four magnetic parts},\ }\href
  {https://doi.org/10.1103/PhysRev.28.334} {\bibfield  {journal} {\bibinfo
  {journal} {Phys. Rev.}\ }\textbf {\bibinfo {volume} {28}},\ \bibinfo {pages}
  {334} (\bibinfo {year} {1926})}\BibitemShut {NoStop}%
\bibitem [{\citenamefont {Curl}\ and\ \citenamefont
  {Kilpatrick}(1960)}]{CURL60}%
  \BibitemOpen
  \bibfield  {author} {\bibinfo {author} {\bibfnamefont {R.~F.}\ \bibnamefont
  {Curl}}\ and\ \bibinfo {author} {\bibfnamefont {J.~E.}\ \bibnamefont
  {Kilpatrick}},\ }\bibfield  {title} {\bibinfo {title} {Atomic term symbols by
  group theory},\ }\href {https://doi.org/10.1119/1.1935804} {\bibfield
  {journal} {\bibinfo  {journal} {Am. J. Phys.}\ }\textbf {\bibinfo {volume}
  {28}},\ \bibinfo {pages} {357} (\bibinfo {year} {1960})}\BibitemShut
  {NoStop}%
\bibitem [{\citenamefont {Karayianis}(1965)}]{KARAYIANIS65}%
  \BibitemOpen
  \bibfield  {author} {\bibinfo {author} {\bibfnamefont {N.}~\bibnamefont
  {Karayianis}},\ }\bibfield  {title} {\bibinfo {title} {Atomic terms for
  equivalent electrons},\ }\href {https://doi.org/doi: 10.1063/1.1704761}
  {\bibfield  {journal} {\bibinfo  {journal} {J. Math. Phys.}\ }\textbf
  {\bibinfo {volume} {6}},\ \bibinfo {pages} {1204} (\bibinfo {year}
  {1965})}\BibitemShut {NoStop}%
\bibitem [{\citenamefont {Katriel}\ and\ \citenamefont
  {Novoselsky}(1989)}]{KATRIEL89}%
  \BibitemOpen
  \bibfield  {author} {\bibinfo {author} {\bibfnamefont {J.}~\bibnamefont
  {Katriel}}\ and\ \bibinfo {author} {\bibfnamefont {A.}~\bibnamefont
  {Novoselsky}},\ }\bibfield  {title} {\bibinfo {title} {Term multiplicities in
  the {LS}-coupling scheme},\ }\href
  {https://doi.org/10.1088/0305-4470/22/9/014} {\bibfield  {journal} {\bibinfo
  {journal} {J. Phys. A: Math. Gen.}\ }\textbf {\bibinfo {volume} {22}},\
  \bibinfo {pages} {1245} (\bibinfo {year} {1989})}\BibitemShut {NoStop}%
\bibitem [{\citenamefont {Xu}\ and\ \citenamefont {Dai}(2006)}]{XU06}%
  \BibitemOpen
  \bibfield  {author} {\bibinfo {author} {\bibfnamefont {R.}~\bibnamefont
  {Xu}}\ and\ \bibinfo {author} {\bibfnamefont {Z.}~\bibnamefont {Dai}},\
  }\bibfield  {title} {\bibinfo {title} {Alternative mathematical technique to
  determine {LS} spectral terms},\ }\href
  {https://doi.org/10.1088/0953-4075/39/16/007} {\bibfield  {journal} {\bibinfo
   {journal} {J. Phys. B: At. Mol. Opt. Phys.}\ }\textbf {\bibinfo {volume}
  {39}},\ \bibinfo {pages} {3221} (\bibinfo {year} {2006})}\BibitemShut
  {NoStop}%
\bibitem [{\citenamefont {Bauche}\ \emph {et~al.}(1988)\citenamefont {Bauche},
  \citenamefont {Bauche-Arnoult},\ and\ \citenamefont {Klapisch}}]{BAUCHE88}%
  \BibitemOpen
  \bibfield  {author} {\bibinfo {author} {\bibfnamefont {J.}~\bibnamefont
  {Bauche}}, \bibinfo {author} {\bibfnamefont {C.}~\bibnamefont
  {Bauche-Arnoult}},\ and\ \bibinfo {author} {\bibfnamefont {M.}~\bibnamefont
  {Klapisch}},\ }\bibfield  {title} {\bibinfo {title} {Transition arrays in the
  spectra of ionized atoms},\ }\href
  {https://doi.org/10.1016/S0065-2199(08)60107-4} {\bibfield  {journal}
  {\bibinfo  {journal} {Adv. At. Mol. Phys.}\ }\textbf {\bibinfo {volume}
  {23}},\ \bibinfo {pages} {131} (\bibinfo {year} {1988})}\BibitemShut
  {NoStop}%
\bibitem [{\citenamefont {Moszkowski}(1960)}]{MOSZKOWSKI60}%
  \BibitemOpen
  \bibfield  {author} {\bibinfo {author} {\bibfnamefont {S.~A.}\ \bibnamefont
  {Moszkowski}},\ }\href {https://www.osti.gov/biblio/4097335} {\emph {\bibinfo
  {title} {Some statistical properties of level and line distributions in
  atomic spectra}}},\ \bibinfo {type} {Tech. Rep.}\ (\bibinfo  {institution}
  {RAND Corp., Santa Monica, California},\ \bibinfo {year} {1960})\BibitemShut
  {NoStop}%
\bibitem [{\citenamefont {Bancewicz}\ and\ \citenamefont
  {Karwowski}(1984)}]{BANCEWICZ84}%
  \BibitemOpen
  \bibfield  {author} {\bibinfo {author} {\bibfnamefont {M.}~\bibnamefont
  {Bancewicz}}\ and\ \bibinfo {author} {\bibfnamefont {J.}~\bibnamefont
  {Karwowski}},\ }\bibfield  {title} {\bibinfo {title} {A study on atomic
  energy level distribution},\ }\href@noop {} {\bibfield  {journal} {\bibinfo
  {journal} {Acta Phys. Pol. Ser. A}\ }\textbf {\bibinfo {volume} {65}},\
  \bibinfo {pages} {279} (\bibinfo {year} {1984})}\BibitemShut {NoStop}%
\bibitem [{\citenamefont {Bauche}\ and\ \citenamefont
  {Bauche-Arnoult}(1987)}]{BAUCHE87}%
  \BibitemOpen
  \bibfield  {author} {\bibinfo {author} {\bibfnamefont {J.}~\bibnamefont
  {Bauche}}\ and\ \bibinfo {author} {\bibfnamefont {C.}~\bibnamefont
  {Bauche-Arnoult}},\ }\bibfield  {title} {\bibinfo {title} {Level and line
  statistic in atomic spectra},\ }\href
  {https://doi.org/10.1088/0022-3700/20/8/006} {\bibfield  {journal} {\bibinfo
  {journal} {J. Phys. B: At. Mol. Opt. Phys.}\ }\textbf {\bibinfo {volume}
  {20}},\ \bibinfo {pages} {1659} (\bibinfo {year} {1987})}\BibitemShut
  {NoStop}%
\bibitem [{\citenamefont {Bauche}\ and\ \citenamefont
  {Bauche-Arnoult}(1990)}]{BAUCHE90}%
  \BibitemOpen
  \bibfield  {author} {\bibinfo {author} {\bibfnamefont {J.}~\bibnamefont
  {Bauche}}\ and\ \bibinfo {author} {\bibfnamefont {C.}~\bibnamefont
  {Bauche-Arnoult}},\ }\bibfield  {title} {\bibinfo {title} {Statistical
  properties of atomic spectra},\ }\href@noop {} {\bibfield  {journal}
  {\bibinfo  {journal} {Comp. Phys. Rep.}\ }\textbf {\bibinfo {volume} {12}},\
  \bibinfo {pages} {1} (\bibinfo {year} {1990})}\BibitemShut {NoStop}%
\bibitem [{\citenamefont {Gilleron}\ and\ \citenamefont
  {Pain}(2009)}]{GILLERON09}%
  \BibitemOpen
  \bibfield  {author} {\bibinfo {author} {\bibfnamefont {F.}~\bibnamefont
  {Gilleron}}\ and\ \bibinfo {author} {\bibfnamefont {J.-C.}\ \bibnamefont
  {Pain}},\ }\bibfield  {title} {\bibinfo {title} {Efficient methods for
  calculating the number of states, levels and lines in atomic
  configurations},\ }\href
  {https://doi.org/https://doi.org/10.1016/j.hedp.2009.04.007} {\bibfield
  {journal} {\bibinfo  {journal} {High Energy Density Phys.}\ }\textbf
  {\bibinfo {volume} {5}},\ \bibinfo {pages} {320 } (\bibinfo {year}
  {2009})}\BibitemShut {NoStop}%
\bibitem [{\citenamefont {Porcherot}\ \emph {et~al.}(2011)\citenamefont
  {Porcherot}, \citenamefont {Pain}, \citenamefont {Gilleron},\ and\
  \citenamefont {Blenski}}]{PORCHEROT11}%
  \BibitemOpen
  \bibfield  {author} {\bibinfo {author} {\bibfnamefont {Q.}~\bibnamefont
  {Porcherot}}, \bibinfo {author} {\bibfnamefont {J.-C.}\ \bibnamefont {Pain}},
  \bibinfo {author} {\bibfnamefont {F.}~\bibnamefont {Gilleron}},\ and\
  \bibinfo {author} {\bibfnamefont {T.}~\bibnamefont {Blenski}},\ }\bibfield
  {title} {\bibinfo {title} {A consistent approach for mixed detailed and
  statistical calculation of opacities in hot plasmas},\ }\href
  {https://doi.org/10.1016/j.hedp.2011.05.001} {\bibfield  {journal} {\bibinfo
  {journal} {High Energy Density Phys.}\ }\textbf {\bibinfo {volume} {7}},\
  \bibinfo {pages} {234} (\bibinfo {year} {2011})}\BibitemShut {NoStop}%
\bibitem [{\citenamefont {Pain}\ \emph {et~al.}(2012)\citenamefont {Pain},
  \citenamefont {Gilleron}, \citenamefont {Bauche},\ and\ \citenamefont
  {Bauche-Arnoult}}]{PAIN12}%
  \BibitemOpen
  \bibfield  {author} {\bibinfo {author} {\bibfnamefont {J.-C.}\ \bibnamefont
  {Pain}}, \bibinfo {author} {\bibfnamefont {F.}~\bibnamefont {Gilleron}},
  \bibinfo {author} {\bibfnamefont {J.}~\bibnamefont {Bauche}},\ and\ \bibinfo
  {author} {\bibfnamefont {C.}~\bibnamefont {Bauche-Arnoult}},\ }\bibfield
  {title} {\bibinfo {title} {Statistics of electric-quadrupole lines in atomic
  spectra},\ }\href {https://doi.org/10.1088/0953-4075/45/13/135006} {\bibfield
   {journal} {\bibinfo  {journal} {J. Phys. B: At. Mol. Opt. Phys.}\ }\textbf
  {\bibinfo {volume} {45}},\ \bibinfo {pages} {135006} (\bibinfo {year}
  {2012})}\BibitemShut {NoStop}%
\bibitem [{\citenamefont {Bauche}\ and\ \citenamefont
  {Coss{\'e}}(1997)}]{BAUCHE97}%
  \BibitemOpen
  \bibfield  {author} {\bibinfo {author} {\bibfnamefont {J.}~\bibnamefont
  {Bauche}}\ and\ \bibinfo {author} {\bibfnamefont {P.}~\bibnamefont
  {Coss{\'e}}},\ }\bibfield  {title} {\bibinfo {title} {Odd-even staggering in
  the {J} and {L} distributions of atomic configurations},\ }\href
  {https://doi.org/10.1088/0953-4075/30/6/010} {\bibfield  {journal} {\bibinfo
  {journal} {J. Phys. B: At. Mol. Opt. Phys.}\ }\textbf {\bibinfo {volume}
  {30}},\ \bibinfo {pages} {1411} (\bibinfo {year} {1997})}\BibitemShut
  {NoStop}%
\bibitem [{\citenamefont {Pain}(2013)}]{PAIN13}%
  \BibitemOpen
  \bibfield  {author} {\bibinfo {author} {\bibfnamefont {J.-C.}\ \bibnamefont
  {Pain}},\ }\bibfield  {title} {\bibinfo {title} {Regularities and symmetries
  in atomic structure and spectra},\ }\href
  {https://doi.org/https://doi.org/10.1016/j.hedp.2013.04.007} {\bibfield
  {journal} {\bibinfo  {journal} {High Energy Density Phys.}\ }\textbf
  {\bibinfo {volume} {9}},\ \bibinfo {pages} {392} (\bibinfo {year}
  {2013})}\BibitemShut {NoStop}%
\bibitem [{\citenamefont {Moszkowski}(1962)}]{MOSZKOWSKI62}%
  \BibitemOpen
  \bibfield  {author} {\bibinfo {author} {\bibfnamefont {S.~A.}\ \bibnamefont
  {Moszkowski}},\ }\bibfield  {title} {\bibinfo {title} {On the energy
  distribution of terms and line arrays in atomic spectra},\ }\href
  {https://doi.org/10.1143/PTP.28.1} {\bibfield  {journal} {\bibinfo  {journal}
  {Progr. Theor. Phys.}\ }\textbf {\bibinfo {volume} {28}},\ \bibinfo {pages}
  {1} (\bibinfo {year} {1962})}\BibitemShut {NoStop}%
\bibitem [{\citenamefont {Ginocchio}(1973)}]{GINOCCHIO73}%
  \BibitemOpen
  \bibfield  {author} {\bibinfo {author} {\bibfnamefont {J.~N.}\ \bibnamefont
  {Ginocchio}},\ }\bibfield  {title} {\bibinfo {title} {Operator averages in a
  shell-model basis},\ }\href {https://doi.org/10.1103/PhysRevC.8.135}
  {\bibfield  {journal} {\bibinfo  {journal} {Phys. Rev. C}\ }\textbf {\bibinfo
  {volume} {8}},\ \bibinfo {pages} {135} (\bibinfo {year} {1973})}\BibitemShut
  {NoStop}%
\bibitem [{\citenamefont {Karazija}(1991)}]{KARAZIJA91b}%
  \BibitemOpen
  \bibfield  {author} {\bibinfo {author} {\bibfnamefont {R.}~\bibnamefont
  {Karazija}},\ }\bibfield  {title} {\bibinfo {title} {Evaluation of explicit
  expressions for mean characteristics of atomic spectra},\ }\href
  {https://doi.org/10.1007/BF03054151 (broken?)} {\bibfield  {journal}
  {\bibinfo  {journal} {Acta Phys. Hung.}\ }\textbf {\bibinfo {volume} {70}},\
  \bibinfo {pages} {367} (\bibinfo {year} {1991})}\BibitemShut {NoStop}%
\bibitem [{\citenamefont {Ku{\v{c}}as}\ \emph {et~al.}(1995)\citenamefont
  {Ku{\v{c}}as}, \citenamefont {Jonauskas}, \citenamefont {Karazija},\ and\
  \citenamefont {Martinson}}]{KUCAS95II}%
  \BibitemOpen
  \bibfield  {author} {\bibinfo {author} {\bibfnamefont {S.}~\bibnamefont
  {Ku{\v{c}}as}}, \bibinfo {author} {\bibfnamefont {V.}~\bibnamefont
  {Jonauskas}}, \bibinfo {author} {\bibfnamefont {R.}~\bibnamefont
  {Karazija}},\ and\ \bibinfo {author} {\bibfnamefont {I.}~\bibnamefont
  {Martinson}},\ }\bibfield  {title} {\bibinfo {title} {Global characteristics
  of atomic spectra and their use for the analysis of spectra. {II}.
  characteristic emission spectra},\ }\href
  {https://doi.org/10.1088/0031-8949/51/5/004} {\bibfield  {journal} {\bibinfo
  {journal} {Phys. Scr.}\ }\textbf {\bibinfo {volume} {51}},\ \bibinfo {pages}
  {566} (\bibinfo {year} {1995})}\BibitemShut {NoStop}%
\bibitem [{\citenamefont {Gilleron}\ \emph {et~al.}(2008)\citenamefont
  {Gilleron}, \citenamefont {Pain}, \citenamefont {Bauche},\ and\ \citenamefont
  {Bauche-Arnoult}}]{GILLERON08}%
  \BibitemOpen
  \bibfield  {author} {\bibinfo {author} {\bibfnamefont {F.}~\bibnamefont
  {Gilleron}}, \bibinfo {author} {\bibfnamefont {J.~C.}\ \bibnamefont {Pain}},
  \bibinfo {author} {\bibfnamefont {J.}~\bibnamefont {Bauche}},\ and\ \bibinfo
  {author} {\bibfnamefont {C.}~\bibnamefont {Bauche-Arnoult}},\ }\bibfield
  {title} {\bibinfo {title} {Impact of high-order moments on the statistical
  modeling of transition arrays},\ }\href
  {https://doi.org/10.1103/PhysRevE.77.026708} {\bibfield  {journal} {\bibinfo
  {journal} {Phys. Rev. E}\ }\textbf {\bibinfo {volume} {77}},\ \bibinfo
  {pages} {026708} (\bibinfo {year} {2008})}\BibitemShut {NoStop}%
\bibitem [{\citenamefont {Kyni\.ene}\ \emph {et~al.}(2002)\citenamefont
  {Kyni\.ene}, \citenamefont {Karazija},\ and\ \citenamefont
  {Jonauskas}}]{KYNIENE02}%
  \BibitemOpen
  \bibfield  {author} {\bibinfo {author} {\bibfnamefont {A.}~\bibnamefont
  {Kyni\.ene}}, \bibinfo {author} {\bibfnamefont {R.}~\bibnamefont
  {Karazija}},\ and\ \bibinfo {author} {\bibfnamefont {V.}~\bibnamefont
  {Jonauskas}},\ }\bibfield  {title} {\bibinfo {title} {Statistical properties
  of {Auger} amplitudes and rates},\ }\href
  {https://doi.org/10.1016/S0368-2048(01)00356-5} {\bibfield  {journal}
  {\bibinfo  {journal} {J. Electron Spectrosc. Relat. Phenom.}\ }\textbf
  {\bibinfo {volume} {122}},\ \bibinfo {pages} {181} (\bibinfo {year}
  {2002})}\BibitemShut {NoStop}%
\bibitem [{\citenamefont {Pain}\ and\ \citenamefont
  {Poirier}(2020)}]{Pain2020}%
  \BibitemOpen
  \bibfield  {author} {\bibinfo {author} {\bibfnamefont {J.-C.}\ \bibnamefont
  {Pain}}\ and\ \bibinfo {author} {\bibfnamefont {M.}~\bibnamefont {Poirier}},\
  }\bibfield  {title} {\bibinfo {title} {Analytical and numerical expressions
  for the number of atomic configurations contained in a supershell},\ }\href
  {https://doi.org/10.1088/1361-6455/ab81ea} {\bibfield  {journal} {\bibinfo
  {journal} {J. Phys. B: At. Mol. Opt. Phys.}\ }\textbf {\bibinfo {volume}
  {53}},\ \bibinfo {pages} {115002} (\bibinfo {year} {2020})}\BibitemShut
  {NoStop}%
\bibitem [{\citenamefont {Condon}\ and\ \citenamefont
  {Shortley}(1935)}]{Condon1935}%
  \BibitemOpen
  \bibfield  {author} {\bibinfo {author} {\bibfnamefont {E.~U.}\ \bibnamefont
  {Condon}}\ and\ \bibinfo {author} {\bibfnamefont {G.~H.}\ \bibnamefont
  {Shortley}},\ }\href@noop {} {\emph {\bibinfo {title} {The theory of atomic
  spectra}}}\ (\bibinfo  {publisher} {Cambridge University Press},\ \bibinfo
  {address} {Cambridge, UK},\ \bibinfo {year} {1935})\BibitemShut {NoStop}%
\bibitem [{\citenamefont {Stuart}\ and\ \citenamefont
  {Ord}(1994)}]{Stuart1994}%
  \BibitemOpen
  \bibfield  {author} {\bibinfo {author} {\bibfnamefont {A.}~\bibnamefont
  {Stuart}}\ and\ \bibinfo {author} {\bibfnamefont {J.~K.}\ \bibnamefont
  {Ord}},\ }\href@noop {} {\emph {\bibinfo {title} {Kendall's Advanced Theory
  of Statistics -- Distribution Theory}}},\ Vol.~\bibinfo {volume} {1}\
  (\bibinfo  {publisher} {John Wiley and Sons},\ \bibinfo {address} {London
  UK},\ \bibinfo {year} {1994})\BibitemShut {NoStop}%
\bibitem [{\citenamefont {Talmi}(2005)}]{Talmi2005}%
  \BibitemOpen
  \bibfield  {author} {\bibinfo {author} {\bibfnamefont {I.}~\bibnamefont
  {Talmi}},\ }\bibfield  {title} {\bibinfo {title} {Number of states with given
  spin {$J$} of $n$ fermions in a $j$ orbit},\ }\href
  {https://doi.org/10.1103/PhysRevC.72.037302} {\bibfield  {journal} {\bibinfo
  {journal} {Phys. Rev. C}\ }\textbf {\bibinfo {volume} {72}},\ \bibinfo
  {pages} {037302} (\bibinfo {year} {2005})}\BibitemShut {NoStop}%
\bibitem [{\citenamefont {Andrews}(1984)}]{ANDREWS1984}%
  \BibitemOpen
  \bibfield  {author} {\bibinfo {author} {\bibfnamefont {G.~E.}\ \bibnamefont
  {Andrews}},\ }\href {https://doi.org/10.1017/CBO9780511608650} {\emph
  {\bibinfo {title} {The Theory of Partitions}}},\ Encyclopedia of Mathematics
  and its Applications\ (\bibinfo  {publisher} {Cambridge University Press},\
  \bibinfo {year} {1984})\BibitemShut {NoStop}%
\bibitem [{\citenamefont {Abramowitz}\ and\ \citenamefont
  {Stegun}(1972)}]{Abramowitz1972}%
  \BibitemOpen
  \bibfield  {author} {\bibinfo {author} {\bibfnamefont {M.}~\bibnamefont
  {Abramowitz}}\ and\ \bibinfo {author} {\bibfnamefont {I.}~\bibnamefont
  {Stegun}},\ }\href@noop {} {\emph {\bibinfo {title} {Handbook of Mathematical
  Functions}}}\ (\bibinfo  {publisher} {National Bureau of Standards},\
  \bibinfo {address} {Washington DC, USA},\ \bibinfo {year} {1972})\BibitemShut
  {NoStop}%
\bibitem [{\citenamefont {Ginocchio}\ and\ \citenamefont
  {Yen}(1975)}]{GINOCCHIO1975}%
  \BibitemOpen
  \bibfield  {author} {\bibinfo {author} {\bibfnamefont {J.}~\bibnamefont
  {Ginocchio}}\ and\ \bibinfo {author} {\bibfnamefont {M.}~\bibnamefont
  {Yen}},\ }\bibfield  {title} {\bibinfo {title} {The dependence of shell model
  state densities on angular momentum},\ }\href
  {https://doi.org/10.1016/0375-9474(75)90373-5} {\bibfield  {journal}
  {\bibinfo  {journal} {Nucl. Phys. A}\ }\textbf {\bibinfo {volume} {239}},\
  \bibinfo {pages} {365} (\bibinfo {year} {1975})}\BibitemShut {NoStop}%
\bibitem [{\citenamefont {Shevelko}\ \emph {et~al.}(1998)\citenamefont
  {Shevelko}, \citenamefont {Shmaenok}, \citenamefont {Churilov}, \citenamefont
  {Bastiaensen},\ and\ \citenamefont {Bijkerk}}]{SHEVELKO1998}%
  \BibitemOpen
  \bibfield  {author} {\bibinfo {author} {\bibfnamefont {A.~P.}\ \bibnamefont
  {Shevelko}}, \bibinfo {author} {\bibfnamefont {L.~A.}\ \bibnamefont
  {Shmaenok}}, \bibinfo {author} {\bibfnamefont {S.~S.}\ \bibnamefont
  {Churilov}}, \bibinfo {author} {\bibfnamefont {R.~K. F.~J.}\ \bibnamefont
  {Bastiaensen}},\ and\ \bibinfo {author} {\bibfnamefont {F.}~\bibnamefont
  {Bijkerk}},\ }\bibfield  {title} {\bibinfo {title} {Extreme ultraviolet
  spectroscopy of a laser plasma source for lithography},\ }\href
  {https://doi.org/10.1088/0031-8949/57/2/023} {\bibfield  {journal} {\bibinfo
  {journal} {Phys. Scr.}\ }\textbf {\bibinfo {volume} {57}},\ \bibinfo {pages}
  {276} (\bibinfo {year} {1998})}\BibitemShut {NoStop}%
\bibitem [{\citenamefont {Hvelplund}\ \emph {et~al.}(1981)\citenamefont
  {Hvelplund}, \citenamefont {Haugen}, \citenamefont {Knudsen}, \citenamefont
  {Andersen}, \citenamefont {Damsgaard},\ and\ \citenamefont
  {Fukusawa}}]{HVELPLUND1981}%
  \BibitemOpen
  \bibfield  {author} {\bibinfo {author} {\bibfnamefont {P.}~\bibnamefont
  {Hvelplund}}, \bibinfo {author} {\bibfnamefont {H.~K.}\ \bibnamefont
  {Haugen}}, \bibinfo {author} {\bibfnamefont {H.}~\bibnamefont {Knudsen}},
  \bibinfo {author} {\bibfnamefont {L.}~\bibnamefont {Andersen}}, \bibinfo
  {author} {\bibfnamefont {H.}~\bibnamefont {Damsgaard}},\ and\ \bibinfo
  {author} {\bibfnamefont {F.}~\bibnamefont {Fukusawa}},\ }\bibfield  {title}
  {\bibinfo {title} {Electron capture into high-lying {R}ydberg states in
  collisions between multiply charged ions and {H2}},\ }\href
  {https://doi.org/10.1088/0031-8949/24/1a/009} {\bibfield  {journal} {\bibinfo
   {journal} {Phys. Scr.}\ }\textbf {\bibinfo {volume} {24}},\ \bibinfo {pages}
  {40} (\bibinfo {year} {1981})}\BibitemShut {NoStop}%
\bibitem [{\citenamefont {Zigler}\ \emph {et~al.}(1987)\citenamefont {Zigler},
  \citenamefont {Givon}, \citenamefont {Yarkoni}, \citenamefont {Kishinevsky},
  \citenamefont {Goldberg}, \citenamefont {Arad},\ and\ \citenamefont
  {Klapisch}}]{ZIGLER1987}%
  \BibitemOpen
  \bibfield  {author} {\bibinfo {author} {\bibfnamefont {A.}~\bibnamefont
  {Zigler}}, \bibinfo {author} {\bibfnamefont {M.}~\bibnamefont {Givon}},
  \bibinfo {author} {\bibfnamefont {E.}~\bibnamefont {Yarkoni}}, \bibinfo
  {author} {\bibfnamefont {M.}~\bibnamefont {Kishinevsky}}, \bibinfo {author}
  {\bibfnamefont {E.}~\bibnamefont {Goldberg}}, \bibinfo {author}
  {\bibfnamefont {B.}~\bibnamefont {Arad}},\ and\ \bibinfo {author}
  {\bibfnamefont {M.}~\bibnamefont {Klapisch}},\ }\bibfield  {title} {\bibinfo
  {title} {Use of unresolved transition arrays for plasma diagnostics},\ }\href
  {https://doi.org/10.1103/PhysRevA.35.280} {\bibfield  {journal} {\bibinfo
  {journal} {Phys. Rev. A}\ }\textbf {\bibinfo {volume} {35}},\ \bibinfo
  {pages} {280} (\bibinfo {year} {1987})}\BibitemShut {NoStop}%
\bibitem [{\citenamefont {Radtke}\ \emph {et~al.}(2001)\citenamefont {Radtke},
  \citenamefont {Biedermann}, \citenamefont {Schwob}, \citenamefont
  {Mandelbaum},\ and\ \citenamefont {Doron}}]{RADTKE2001}%
  \BibitemOpen
  \bibfield  {author} {\bibinfo {author} {\bibfnamefont {R.}~\bibnamefont
  {Radtke}}, \bibinfo {author} {\bibfnamefont {C.}~\bibnamefont {Biedermann}},
  \bibinfo {author} {\bibfnamefont {J.~L.}\ \bibnamefont {Schwob}}, \bibinfo
  {author} {\bibfnamefont {P.}~\bibnamefont {Mandelbaum}},\ and\ \bibinfo
  {author} {\bibfnamefont {R.}~\bibnamefont {Doron}},\ }\bibfield  {title}
  {\bibinfo {title} {Line and band emission from tungsten ions with charge
  $21+$ to $45+$ in the {45--70-\AA} range},\ }\href
  {https://doi.org/10.1103/PhysRevA.64.012720} {\bibfield  {journal} {\bibinfo
  {journal} {Phys. Rev. A}\ }\textbf {\bibinfo {volume} {64}},\ \bibinfo
  {pages} {012720} (\bibinfo {year} {2001})}\BibitemShut {NoStop}%
\bibitem [{\citenamefont {Jonauskas}\ \emph {et~al.}(2012)\citenamefont
  {Jonauskas}, \citenamefont {Kynien{\.{e}}}, \citenamefont {Kisielius},\ and\
  \citenamefont {Masys}}]{Jonauskas2012}%
  \BibitemOpen
  \bibfield  {author} {\bibinfo {author} {\bibfnamefont {V.}~\bibnamefont
  {Jonauskas}}, \bibinfo {author} {\bibfnamefont {A.}~\bibnamefont
  {Kynien{\.{e}}}}, \bibinfo {author} {\bibfnamefont {R.}~\bibnamefont
  {Kisielius}},\ and\ \bibinfo {author} {\bibfnamefont {{\v{S}}.}~\bibnamefont
  {Masys}},\ }\bibfield  {title} {\bibinfo {title} {Theoretical study of
  {W$^{20+}$} spectra formation in {EBIT} plasma},\ }\href
  {https://doi.org/10.1088/1742-6596/388/4/042016} {\bibfield  {journal}
  {\bibinfo  {journal} {J. Phys. Conf. Ser.}\ }\textbf {\bibinfo {volume}
  {388}},\ \bibinfo {pages} {042016} (\bibinfo {year} {2012})}\BibitemShut
  {NoStop}%
\bibitem [{\citenamefont {O'Sullivan}\ \emph {et~al.}(2015)\citenamefont
  {O'Sullivan}, \citenamefont {Dunne}, \citenamefont {Higashiguchi},
  \citenamefont {Li}, \citenamefont {Liu}, \citenamefont {Lokasani},
  \citenamefont {Long}, \citenamefont {Ohashi}, \citenamefont {O'Reilly},
  \citenamefont {Sheridan}, \citenamefont {Sokell}, \citenamefont {Suzuki},\
  and\ \citenamefont {Wu}}]{OSullivan2015}%
  \BibitemOpen
  \bibfield  {author} {\bibinfo {author} {\bibfnamefont {G.}~\bibnamefont
  {O'Sullivan}}, \bibinfo {author} {\bibfnamefont {P.}~\bibnamefont {Dunne}},
  \bibinfo {author} {\bibfnamefont {T.}~\bibnamefont {Higashiguchi}}, \bibinfo
  {author} {\bibfnamefont {B.}~\bibnamefont {Li}}, \bibinfo {author}
  {\bibfnamefont {L.}~\bibnamefont {Liu}}, \bibinfo {author} {\bibfnamefont
  {R.}~\bibnamefont {Lokasani}}, \bibinfo {author} {\bibfnamefont
  {E.}~\bibnamefont {Long}}, \bibinfo {author} {\bibfnamefont {H.}~\bibnamefont
  {Ohashi}}, \bibinfo {author} {\bibfnamefont {F.}~\bibnamefont {O'Reilly}},
  \bibinfo {author} {\bibfnamefont {P.}~\bibnamefont {Sheridan}}, \bibinfo
  {author} {\bibfnamefont {E.}~\bibnamefont {Sokell}}, \bibinfo {author}
  {\bibfnamefont {C.}~\bibnamefont {Suzuki}},\ and\ \bibinfo {author}
  {\bibfnamefont {T.}~\bibnamefont {Wu}},\ }\bibfield  {title} {\bibinfo
  {title} {Spectroscopy for identification of plasma sources for lithography
  and water window imaging},\ }\href
  {https://doi.org/10.1088/1742-6596/635/1/012026} {\bibfield  {journal}
  {\bibinfo  {journal} {J. Phys. Conf. Ser.}\ }\textbf {\bibinfo {volume}
  {635}},\ \bibinfo {pages} {012026} (\bibinfo {year} {2015})}\BibitemShut
  {NoStop}%
\bibitem [{\citenamefont {Blinnikov}\ and\ \citenamefont
  {Moessner}(1998)}]{Blinnikov1998}%
  \BibitemOpen
  \bibfield  {author} {\bibinfo {author} {\bibfnamefont {S.}~\bibnamefont
  {Blinnikov}}\ and\ \bibinfo {author} {\bibfnamefont {R.}~\bibnamefont
  {Moessner}},\ }\bibfield  {title} {\bibinfo {title} {Expansions for nearly
  {G}aussian distributions},\ }\href {https://doi.org/10.1051/aas:1998221}
  {\bibfield  {journal} {\bibinfo  {journal} {Astron. Astrophys. Suppl. Ser.}\
  }\textbf {\bibinfo {volume} {130}},\ \bibinfo {pages} {193} (\bibinfo {year}
  {1998})}\BibitemShut {NoStop}%
\bibitem [{\citenamefont {de~Kock}\ \emph {et~al.}(2011)\citenamefont
  {de~Kock}, \citenamefont {Eggers},\ and\ \citenamefont
  {Schmiegel}}]{deKock2011}%
  \BibitemOpen
  \bibfield  {author} {\bibinfo {author} {\bibfnamefont {M.~B.}\ \bibnamefont
  {de~Kock}}, \bibinfo {author} {\bibfnamefont {H.~C.}\ \bibnamefont
  {Eggers}},\ and\ \bibinfo {author} {\bibfnamefont {J.}~\bibnamefont
  {Schmiegel}},\ }\bibfield  {title} {\bibinfo {title} {Edgeworth versus
  {G}ram-{C}harlier series: x-cumulant and probability density tests},\ }\href
  {https://doi.org/10.1134/S1547477111090159} {\bibfield  {journal} {\bibinfo
  {journal} {Phys. Part. Nucl. Lett.}\ }\textbf {\bibinfo {volume} {8}},\
  \bibinfo {pages} {1023} (\bibinfo {year} {2011})}\BibitemShut {NoStop}%
\end{thebibliography}%
\end{document}